%% file: paper.tex
\documentclass[prd,aps,superscriptaddress,showpacs,preprint,preprintnumbers,amsmath,amssymb]{revtex4}
\usepackage{graphicx}
\usepackage{dcolumn}
\usepackage{bm}

\usepackage{color}
\usepackage{ulem}
\definecolor{DeepBlue}{rgb}{0,0,0.5}
\definecolor{DeepGreen}{rgb}{0,0.5,0}
\definecolor{DeepRed}{rgb}{0.5,0,0}

\input{template}
\begin{document}
\bibliographystyle{apsrev}



\newcounter{Outline}
\setcounter{Outline}{0}

\newcounter{Intro}
\setcounter{Intro}{1}

\newcounter{General}
\setcounter{General}{1}

\newcounter{NPR}
\setcounter{NPR}{1}

\newcounter{ME}
\setcounter{ME}{1}

\newcounter{Dynamical}
\setcounter{Dynamical}{1}

\newcounter{Conclusions}
\setcounter{Conclusions}{1}

\newcounter{Acknowledgments}
\setcounter{Acknowledgments}{1}

\newcounter{Appendix}
\setcounter{Appendix}{1}

%


\preprint{BNL-HET--6/5, RBRC-606}

\title{Proton decay matrix elements with domain-wall fermions} 

\author{Y.~Aoki}
\affiliation{Physics Department, University of Wuppertal, Gaussstr.~20, 42119
  Wuppertal, Germany}

\author{C.~Dawson}
\affiliation{RIKEN-BNL Research Center, Brookhaven National Laboratory, Upton, NY 11973}

\author{J.~Noaki}
\affiliation{School of Physics and Astronomy, University of Southampton,
Southampton, SO17 1BJ, U.K.}

\author{A.~Soni}
\affiliation{Physics Department, Brookhaven National Laboratory, Upton, NY 11973}

\date{\today}

\begin{abstract}
\input{text_sections/abstract.tex}
\end{abstract}

\maketitle

\newpage


\section{Introduction}
\label{sec:intro}

\ifnum\theIntro=1
\input{text_sections/intro.tex}
\fi


\section{General Description}
\label{sec:general}

\ifnum\theGeneral=1
\input{text_sections/general.tex}
\fi


\section{Operator renormalization}
\label{sec:NPR}

\ifnum\theNPR=1
\input{text_sections/npr.tex}
\fi


\section{Matrix elements}
\label{sec:ME}

\ifnum\theME=1
\input{text_sections/me.tex}
\fi


\section{Dynamical quark effects}
\label{sec:dynamical}

\ifnum\theDynamical=1
\input{text_sections/dynamical.tex}
\fi


\section{Conclusions}
\label{sec:conclusions}

\ifnum\theConclusions=1
\input{text_sections/conclusions.tex}

\fi


\ifnum\theAcknowledgments=1
\begin{acknowledgments}
We thank N.~Christ, B.~Mawhinney for the continuous
support of this study. Discussions with them as well as
with T.~Blum, T.~Izubuchi, K.~Orginos, S.~Ohta and N.~Yamada are
gratefully acknowledged. We thank T.~Izubuchi, again, for checking the
calculation of the low energy constants of proton decay for the
dynamical-quark simulation.
The calculations reported here were done on the 400 Gflops QCDSP
computer \cite{Chen:1998cg} at Columbia University and the 600 Gflops
QCDSP computer \cite{Mawhinney:2000fx} at the RIKEN BNL Research
Center. We thank RIKEN, Brookhaven National Laboratory and the U.S.\
Department of Energy for providing the facilities essential for the
completion of this work.
This research was supported in part by the DOE under grant \#
DE-AC02-98CH10886 (Soni).
\end{acknowledgments}
\fi


\appendix

\ifnum\theAppendix=1
\section{Fields and convention}
\label{sec:app notation}
\input{text_sections/appendix_notation.tex}

\section{Chiral perturbation}
\label{sec:app chpt}
\input{text_sections/appendix_chpt.tex}

\section{Perturbative matching factor}
\label{sec:app pr}
\input{text_sections/appendix_pr.tex}

\section{Numerical tables}
\label{sec:app table}
\input{text_sections/appendix_table.tex}
\fi


\bibliography{paper}

\pagebreak

\end{document}

%% file: template.tex
\newcommand{\psl}{{p\hspace{-5pt}{/}}}
\newcommand{\qsl}{{q\hspace{-5pt}{/}}}
\newcommand{\ksl}{{k\hspace{-5pt}{/}}}
\newcommand{\ndop}{{\mathcal O}^{R/L\;L}}
\newcommand{\msbar}{\overline{\mbox{MS}}}
\newcommand{\Prt}[1]{{\mathcal P}^{#1}} 
\newcommand{\Swt}[1]{{\mathcal S}^{#1}} 

\newcommand{\MSbar}{\overline{MS}}

\def\mpi2{m_\pi^2}
\def\mK2{m_K^2}

\def\mres{m_{\rm res}}
\newcommand{\Dslash}{\rlap{/}\kern-2.0pt D}

\newcommand{\bea}{\begin{eqnarray}}
\newcommand{\eea}{\end{eqnarray}}
\newcommand{\be}{\begin{equation}}
\newcommand{\ee}{\end{equation}}

\newenvironment{outline}{
\noindent \framebox{Begin \ outline \hspace{5.0in} }
\begin{enumerate}}
{\end{enumerate}
\noindent \framebox{End \ outline \hspace{5.0in} }
\vspace{0.25in}}

\def\simge{
    \mathrel{\rlap{\raise 0.511ex
        \hbox{$>$}}{\lower 0.511ex \hbox{$\sim$}}}}
\def\simle{
    \mathrel{\rlap{\raise 0.511ex
        \hbox{$<$}}{\lower 0.511ex \hbox{$\sim$}}}}

%% file: text_sections/abstract.tex
%
%

Hadronic matrix elements 
of operators relevant to nucleon decay in 
grand unified theories 
are calculated numerically using lattice QCD.
In this context, the domain-wall fermion formulation, combined with 
non-perturbative renormalization, is used for the
first time. These techniques bring reduction of a large fraction of
the systematic error from the finite lattice spacing. 
Our main effort is devoted to a calculation performed in the quenched
approximation, where the direct 
calculation of the nucleon to pseudoscalar matrix elements, as
well as the indirect estimate of them from the nucleon to vacuum
matrix elements, are performed.
First results, using two flavors of dynamical domain-wall quarks
for the nucleon to vacuum matrix elements are also presented
to address the systematic error of quenching, which appears to be
small compared to the other errors.
Our results suggest that the representative value for the low energy
constants from the nucleon to vacuum matrix elements are given
as $|\alpha|\simeq |\beta|\simeq 0.01$ GeV$^3$.
For a more reliable estimate of the physical low energy matrix elements,
it is better to use the relevant form factors
calculated in the direct method.
The direct method tends to give smaller value of the
form factors, compared to the indirect one, thus enhancing the proton
life-time; indeed for the 
$\pi^0$ final state the difference
between the two methods is quite appreciable.

%% file: text_sections/intro.tex
%
%

\ifnum\theOutline=1
\begin{outline}
\item Nucleon decay is natural for theory side
\item matrix element is important for the estimate of lifetime of proton
\item phenomenological calculations vary factor 10 between smallest to largest
\item old lattice calculations as well
\item recent calc by JLQCD and CP-PACS
\item we do DWF and NPR
\item 1st dynamical simulation result is presented
\end{outline}
\fi

In the standard model baryon number is conserved up to a very good
approximation. 
While it is broken (only) by the electro-weak anomaly, the size of breaking
is extremely small so that baryon number violation
would be extremely difficult to observe in any high energy experiment.
The conservation of baryon number happens to be  exact in the classical
level in the standard model. However, it is not
a consequence of the underlying fundamental symmetries, e.~g.\ gauge 
and Lorentz symmetries. In general, baryon number is not
protected once we extend the gauge groups of the standard model
to larger symmetry groups.
The existence of the baryon number
violating interaction makes protons unstable.
This particular phenomena,  proton decay,
or in general, nucleon decay is one of the most decisive
signals that prove 
that nature has larger fundamental symmetry than that of the standard model.
Grand unified theories (GUTs) 
\cite{Pati:1973XX,Georgi:1974sy},
with or without supersymmetry, possess such 
a feature.

Although proton decay has been searched for in the deep mine
experiments for more than a decade, it has not been observed
\cite{Kobayashi:2005pe}.
The lower bound of the proton lifetime from the experiments
has provided an important stringent constraint on the GUT models.
In fact, the simplest minimal SUSY GUT with $SU(5)$ gauge group is
almost surely excluded \cite{Murayama:2001ur}
\footnote{Note that one can still save the theory by appropriate changes
   in the flavor structure \cite{Bajc:2002bv,Emmanuel-Costa:2003pu}.}.

At low energies, the processes
that allow nucleon decay may be represented in terms of a low energy effective
Hamiltonian made up of Standard Model fields. This effective Hamiltonian
will be dominated by the operators of lowest dimension.
The requirement that the low energy effective theory has 
the same symmetry as the standard model is strong enough to constrain the
form of the possible operators in the effective Hamiltonian
\cite{Weinberg:1979sa,Wilczek:1979hc,Abbott:1980zj}.
The operators consist of three quarks and one lepton
fields as $(\overline{q}^cq)(\overline{l}^cq)$, which lead to
a decay of a nucleon to a pseudoscalar and an anti-lepton.
As the low energy hadrons are involved in the decay, this is a
highly non-perturbative process, and so lattice QCD provides
an ideal tool to analyze it.

Since difficult to calculate the hadronic matrix element
$\langle PS|(\overline{q}^cq)q|N\rangle$ of the three-quark operator
with initial nucleon ($N$) and final pseudoscalar ($PS$) states,
various QCD model approaches were used to  
calculate nucleon to vacuum matrix element
$\langle 0|(\overline{q}^cq)q|N\rangle$,
which in turn gives $\langle PS|(\overline{q}^cq)q|N\rangle$ with the help of 
chiral perturbation theory. In 80's many calculations of
the matrix elements 
\cite{Ioffe:1981kw,Tomozawa:1981rc,Donoghue:1982jm,Meljanac:1982xd,
Krasnikov:1982gf,Thomas:1983ch,Ioffe:1984ju,Brodsky:1984st}
were reported.
These calculations varied by a factor of $O(10)$ between smallest to largest estimate
(see \cite{Brodsky:1984st} and Table \ref{tab:summary_chL}).
As partial width of the decay is proportional to the matrix elements
squared, different model calculations lead to a factor of $O(100)$
difference in proton lifetime. 

The first calculations \cite{Hara:1986hk,Bowler:1988us,Gavela:1989cp}
using lattice QCD were performed in late 80's. These calculations
also took the approach of calculating the nucleon to vacuum matrix 
elements, using chiral perturbation theory to infer the final result.
The nucleon to vacuum matrix elements obtained in these studies disagreed
as well, with a factor of about 5 difference between the smallest
\cite{Gavela:1989cp} to largest \cite{Hara:1986hk} estimate.

There are various sources of systematic errors involved in such
calculations, one of which is the large range extrapolation of 
the matrix elements in the quark mass. As the matrix elements are 
used as the low energy constants of the
chiral Lagrangian, they must be obtained in the chiral limit.
These original lattice studies
necessarily used unphysically
heavy values of the up and down quark masses. In fact, the values
used were typically of the order of the mass of the physical strange quark.

Another source of error comes from measuring the lattice scale. 
Since the low energy constants measured have mass dimension three, their
error -- when quoted in physical units -- receives a contribution equal
to three times the error on this scale. There are various ways of extracting
the lattice scale which need not agree at finite lattice spacing, or
when working in the quenched approximation. Moreover, some
conventional ways to set the lattice scale such as using the $\rho$ meson or
the nucleon mass as an input again, require
a large extrapolation in the mass to reach the physical point.
It is important to note that all
the calculations performed up to now used unimproved Wilson
fermions; a formulation known to have large scaling violations.
It is not surprising that a moderate difference in the 
lattice cut off leads to a large difference in the
low energy constants.

Operator renormalization is another source of systematic
error. The operators which mediate the nucleon decay
must be renormalized in some renormalization 
scheme. This must be the same renormalization scheme that was 
used for the calculation of the Wilson coefficients for the effective 
Hamiltonian, and is usually $\msbar$ scheme.
The common way to perform this calculation 
has been to use lattice perturbation theory,
which has poor convergence property, 
leading to a large systematic error.

There is one important systematics error, which neither the
model or lattice calculations in the 80's address:
even if one gets the correct value of the low energy constants,
the matrix element for the decay could have an appreciable error.
The reason is that the pion resulting from the decay has,
in the center of mass frame, 
a momentum around half of the nucleon mass, where the leading order
chiral expansion may not be a good approximation.

The, more recent, JLQCD work \cite{Aoki:1999tw} was better 
at addressing many of these systematics compared to the old results:
they employed smaller quark masses, and used 
an improved lattice perturbation theory.
Most importantly, the first reliable direct measurement for
the nucleon to pseudoscalar matrix elements were performed.
However, the 
calculation was still performed in the quenched approximation, and
made use of the unimproved Wilson action.
Later, a joint collaboration including some of the original
JLQCD members addressed the issue of the scaling violations
in the restricted case of the 
nucleon to vacuum matrix elements \cite{Tsutsui:2004qc}. It appeared the
value that JLQCD got at a finite lattice spacing was larger by almost
factor 2 than that in the continuum limit, showing the particular
difficulty in taking the continuum limit with the unimproved Wilson fermions.

In this paper we calculate the nucleon decay matrix elements in a 
approach similar to JLQCD, but, with three key differences:
we use domain-wall fermions,
which preserve chiral symmetry to a very high accuracy
(and so are expected to have much reduced scaling violations),
we use a non-perturbative renormalization scheme,
and we also investigate the effects 
of moving away from the quenched approximation.
As mentioned previously, 
all the works which have been done so far for nucleon decay used
the quenched approximation wherein the quark loops
in the propagation of the gluons are neglected. Although
it has been shown that light flavored hadron spectrum is reproduced
by the quenched approximation within 10\% difference of experiment
\cite{Aoki:1999yr}, the systematic error due to this approximation is
process dependent and,
in general, uncontrollable. 
We present the first study of quenching error by calculating the nucleon
to vacuum matrix elements using unquenched $u$, $d$ quarks
and comparing to the quenched result.
As a result of these differences an appreciable reduction 
of the systematic error is expected.

This paper is organized as follows: in Sect.~\ref{sec:general} we
summarize the general properties of the baryon number violating operator
and the nucleon decay matrix elements. Sect.~\ref{sec:NPR} discusses
operator property in view of the renormalization and gives the
detail of how to apply the non-perturbative renormalization scheme
to the nucleon decay operator. The calculation of nucleon decay matrix
elements with the quenched approximation is shown in Sec.~\ref{sec:ME}.
The estimate of the nucleon to vacuum matrix elements with
unquenching $u$, $d$ quark is given in Sec.~\ref{sec:dynamical}.
Sec.~\ref{sec:conclusions} is devoted to the conclusions.
Some results obtained in earlier stages of this work 
have been reported in Refs.~\cite{Aoki:2002ji,Aoki:2003vz,Aoki:2004xe}.

We use the Euclidean lattice formulation, so the 
metric and gamma matrices should be taken as Euclidean.
Dimensional quantities are often written
in the lattice units ($a$). We avoid writing ``$a$'' 
explicitly in most cases.

%% file: text_sections/general.tex
%
%

\ifnum\theOutline=1
\begin{outline}
\item Operator
\item Matrix element
\item Form factor
\item Chiral Lagrangian
\item Lattice calculation set up (details are described in each section)
\end{outline}
\fi

In this section we describe the operators and matrix elements
to calculate and their theoretical background.
Lattice calculation setup is overviewed, which in the later
sections will be discussed in detail.

\subsection{Properties of operators and matrix elements}

The general type of the nucleon decay operator that appears in the low
energy effective Hamiltonian is constrained 
by the symmetries of the standard model, 
$SU(3)_c \times SU(2)_L \times U(1)_Y$ \cite{Weinberg:1979sa,Wilczek:1979hc}.
The quark part must contain three quark fields, as the $SU(3)_c$
singlet made of $3\times 3\times 3$ is the lowest dimensional 
baryon number violating operator which is invariant under $SU(3)_c$.
Another fermion field (lepton or antilepton) is necessary to make the
operator Lorentz invariant. The $SU(2)_L \times U(1)_Y$ symmetry
rules out the possibility of having the antilepton field in the operator,
and restricts the flavor/chirality combination to four types
\cite{Abbott:1980zj}.
In the notation of Weinberg \cite{Weinberg:1979sa}; 
Abbott and Wise \cite{Abbott:1980zj} these
operators read
\footnote{Abbott and Wise reduced Weinberg's six operators to these
four, using (anti) symmetric property under flavor exchange.},
\begin{eqnarray}
 O_{abcd}^{(1)} & = & (D_a^{i}, U_b^j)_R
  (q_{c}^{k \alpha}, l_{d}^{\beta} )_L \epsilon^{ijk}\epsilon^{\alpha\beta}\\
 O_{abcd}^{(2)} & = & (q_a^{i \alpha}, q_b^{j \beta})_L
  (U_{c}^{k}, l_{d})_R \epsilon^{ijk}\epsilon^{\alpha\beta}\\
 \tilde{O}_{abcd}^{(4)} & = & (q_a^{i \alpha}, q_b^{j \beta})_L
  (q_{c}^{k \gamma}, l_{d}^{\delta})_L \epsilon^{ijk}
  \epsilon^{\alpha\beta}\epsilon^{\gamma\delta}\\
 O_{abcd}^{(5)} & = & (D_a^{i}, U_b^j)_R
  (U_{c}^{k}, l_{d})_R \epsilon^{ijk},
\end{eqnarray}
where $l$ is the generic lepton field, $q$ is the left handed quark field,
 $U$ and $D$ denote up and down type right handed
quarks. $a,b,c,d$ are generation numbers,
$i,j,k$ are $SU(3)$ color labels, $\alpha,\beta,\gamma,\delta$ are
$SU(2)$ indices. The inner product $(x,y)_{R/L}$ is defined
as $(x,y)_{R/L}=x^T CP_{R/L} y$, where $C$ is the charge conjugation matrix
and $P_{R/L}$ is the right/left handed projection matrix.
The vector and tensor Dirac matrices have been eliminated in the
expression by Fierz rearrangement.
These dimension six operators 
\footnote{We call all the $qqql$ operators as dimension six.
The naming is just from the naive counting of the field dimension and
nothing to do with the origin of the operator.}
are the lowest dimensional operators that
appear in the low energy effective Hamiltonian written in terms
of Standard Model particles.
Higher dimensional operators are suppressed by inverse powers of the heavy
mass scale characteristic of the fundamental high energy 
theory (eg.~$M_X$ for GUTs).
As is evident from the form of the operator, it breaks baryon 
number ($B$), but
preserves baryon minus lepton number $(B-L)$, leading to 
a decay of nucleon to a pseudoscalar and an antilepton.

The general form of the three quark part of the operator, which
transforms as a spinor, is
\begin{equation}
 {\mathcal O}^{\Gamma\Gamma'}_{uds}= (ud)_{\Gamma} s_{\Gamma'} = \epsilon^{ijk}
  (u^{i\;T} C P_{\Gamma} d^j) P_{\Gamma'} s^k,
  \label{eq:operator}
\end{equation}
where $\Gamma(\Gamma')$ can be either $R$ or $L$.
$u$, $d$, and $s$ are quark fields not necessarily labeling the
real $u$, $d$, $s$ flavors.
From the form of the operator it is evident that
the proton or neutron cannot decay to a final state 
that has strangeness $S<0$. Thus, processes such as
$p\to\overline{K^0}+l^+$  are disallowed. 
The trivial constraint that the mass of the pseudoscalar (PS)
is below that of the nucleon (N), $m_{PS} < m_N$, requires the final
state to be one of 
$\pi^0$, $\pi^\pm$, $K^0$, $K^+$ or $\eta$. Hence the real
physical flavor of a quark in Eq.~(\ref{eq:operator}) is one of 
the three lightest, $u$, $d$, $s$.

We calculate the hadronic matrix element with the  nucleon
and the allowed pseudoscalar states,
\begin{equation}
 \langle PS;\vec{p} | {\mathcal O}^{\Gamma\Gamma'} | N; \vec{k},s \rangle,
\end{equation}
where $\vec{p}$ and $\vec{k}$ are spatial momenta of pseudoscalar and
nucleon respectively, and $s=\pm 1/2$ is the spin of the nucleon. 
Parity transformation yields a relation between  different chirality matrix
elements:
\begin{equation}
 \langle PS;\vec{p} | \ndop | N; \vec{k},s \rangle
  = \gamma_4 \langle PS;-\vec{p} | {\mathcal O}^{L/R\;R} | N;
  -\vec{k},s \rangle. 
\label{eq:me_parity}
\end{equation}
For the range of accuracy that is expected from our calculation,
it is sufficient to assume isospin
symmetry, which further reduces the number of independent matrix
elements. Following Ref.~\cite{Aoki:1999tw} here we list 14
matrix elements and their ``isospin partners''  obtained 
by exchanging $u$ and $d$:
\begin{eqnarray}
 \langle \pi^0 | (ud)_{R/L} u_L | p \rangle &=&
  \langle \pi^0 | (du)_{R/L} d_L | n \rangle,
  \label{eqn:p2pi0}\\
 \langle \pi^+ | (ud)_{R/L} d_L | p \rangle &=&
  -\langle \pi^- | (du)_{R/L} u_L | n \rangle,
  \label{eqn:p2pi+}\\
 \langle  K^0  | (us)_{R/L} u_L | p \rangle &=&
  -\langle  K^+  | (ds)_{R/L} d_L | n \rangle,
 \label{eqn:p2K0a}\\
 \langle  K^+  | (us)_{R/L} d_L | p \rangle &=&
  -\langle  K^0  | (ds)_{R/L} u_L | n \rangle,\\
 \langle  K^+  | (ud)_{R/L} s_L | p \rangle &=&
  -\langle  K^0  | (du)_{R/L} s_L | n \rangle,\\
 \langle  K^+  | (ds)_{R/L} u_L | p \rangle &=&
  -\langle  K^0  | (us)_{R/L} d_L | n \rangle,\\
 \langle \eta  | (ud)_{R/L} u_L | p \rangle &=&
  -\langle \eta  | (du)_{R/L} d_L | n \rangle.
  \label{eqn:p2eta}
\end{eqnarray}
Our interpolating field for each hadron state is 
summarized in appendix \ref{sec:app notation}. The negative
signs on the {\it rhs}
of Eqs.~(\ref{eqn:p2pi0})--(\ref{eqn:p2eta})
appear from the transformation of the interpolating fields 
$J_{\pi^0}\to -J_{\pi^0}$ and $\overline{J}_p\to -\overline{J}_n$ by
interchanging $u$ and $d$. 
There is a relation between final $\pi^0$ and final $\pi^\pm$ matrix
elements in the isospin limit
\begin{equation}
 \langle \pi^+ | (ud)_{R/L} d_L | p \rangle =
  \sqrt{2} \langle \pi^0 | (ud)_{R/L} u_L | p \rangle.
  \label{eq:p2pi0 isospin}
\end{equation}
We call the {\it lhs} of Eqs.~(\ref{eqn:p2pi0}),
(\ref{eqn:p2K0a})--(\ref{eqn:p2eta}) the principal matrix elements.
In the quenched simulation, we are going to calculate these 12
principal matrix
elements. All the other possible matrix elements are obtained  
from the principal matrix elements by
Eqs.~(\ref{eqn:p2pi0})--(\ref{eqn:p2eta}), Eq.~(\ref{eq:me_parity}) and
Eq.~(\ref{eq:p2pi0 isospin}).
Note, however, flavor $SU(3)$ breaking effect of $\eta$ is
not treated in this paper\footnote{It requires to calculate disconnected
diagram,  which is challenging and demanding in the numerical simulation.}.
A nucleon  to pseudoscalar  decay is 
characterized by its initial
three momentum $\vec{k}$, spin $s=\pm 1/2$ and final momentum $\vec{p}$.
By Lorentz covariance
the matrix element is required to have 
the form \cite{Aoki:1999tw},
\begin{equation}
  \langle PS;\vec{p} | \ndop | N;\vec{k},s\rangle = 
  P_{L} [ W_0^{R/L\;L}(q^2) - i \qsl W_q^{R/L\;L}(q^2)] u_N(\vec{k},s),
  \label{eq:formfactor}
\end{equation}
where $u_N$ is the nucleon spinor. The form factors $W_0$ and $W_q$ are 
functions of the 
square of the momentum transfer $q_\mu=k_\mu-p_\mu$. For the physical
decay this is the momentum of the lepton. 
However, on the lattice we work with unphysical
values of the masses and momentum transfer, then extrapolate to the
physical point.
In this case, $W_0$ and $W_q$ are the functions of
$m_N$, $m_{PS}$ and $q^2$.
For the range of masses and momenta used in our simulation
the two terms in the square brackets seem to be of the same order.
This fact implies that the second term is negligible
in the physical amplitude due to the on-shell condition on the
lepton ($-q^2=m_l^2$)  
\footnote{From the numerical calculation where
we set nucleon at rest, we observe $W_0\sim -iq_0W_q$. Thus, with the
existence of lepton spinor $\overline{v}^c$ at left,
$\overline{v}^c(-i\qsl W_q)=\overline{v}^cm_lW_q\sim 
\overline{v}^cm_l/q_0\cdot W_0$.
Since the $q_0$ is the order of the nucleon mass, $-i\qsl W_q$ is
smaller than $W_0$ by an factor $\sim m_l/m_N$.}.
Hence, $W_0$ is called relevant and $W_q$ is irrelevant form factor.
As $-q^2=m_l^2\simeq 0$, $W_0(0)$ is the final target of our calculation.
The parity condition Eq.~(\ref{eq:me_parity}) implies the relation between
form factors with different chirality as
\begin{equation}
 W_x^{R/L\;R}(q^2) = W_x^{L/R\;L}(q^2),
  \label{eq:parity_W}
\end{equation}
for $x=0$, $q$. Note that parity holds only after 
the statistical average over lattice gauge field configurations. 
On a single gauge configuration 
the {\it rhs} and {\it lhs} of Eq.~(\ref{eq:parity_W}) generally differ, and
we can take average of them to get better statistics, which is done
in our analysis.

Once the form factor is calculated the 
partial width of the
$p\to PS+\overline{l}$ decay is obtained as
\begin{equation}
 \Gamma(p\to PS+\overline{l}) = \frac{m_p}{32\pi^2}
  \left[1-\left(\frac{m_{PS}}{m_p}\right)^2\right]^2
  \left|\sum_i C^i W_0^i(p\to PS)\right|^2,
  \label{eq:width}
\end{equation}
where $m_p$ ($m_{PS}$) denotes the proton (pseudoscalar) mass,
and we have set the lepton mass to zero. 
$C^i$ are the Wilson coefficients of the dimension-six operator 
in the low energy effective Hamiltonian,
\begin{equation}
 {\mathcal L}_6 = \sum_i C^i[(qq)(ql)]^i
  = - \sum_i C^i\left[\overline{l^c}{\mathcal O}_{qqq}\right]^i.
\end{equation}
The index $i$ distinguishes the type
(flavor and chirality) of the three quark operator ${\mathcal O}_{qqq}$, 
which is one of those in the matrix elements
Eqs.~(\ref{eqn:p2pi0})--(\ref{eqn:p2eta}), as well as the type of lepton.
While $C_i$ and $W_0^i$ depends on the renormalization scheme and scale,
the dependence cancels out in the product.
In this work we use a renormalization scale
$\mu=2$ GeV and $\msbar$ scheme with the naive dimensional 
regularization (NDR) for $W_0^i$.
With $W_0^i$ calculated in this work, and a knowledge of $C_i$, which
depend on the particular GUT model, proton lifetime can be estimated 
through Eq.~(\ref{eq:width}).

\subsection{Usage of chiral perturbation theory}

In this work we discuss the calculation of the
form factors for the principal matrix
elements  by means of numerical lattice simulations. 
To obtain the results at the physical kinematics in terms of quark
masses and pseudoscalar momentum, it is necessary to know the dependence of the
form factors on these parameters, especially for the $u$, $d$ quark masses
as these are not 
attainable on the lattice with the present 
algorithmic and computational resources.
Chiral perturbation theory ($\chi$PT) gives such an information.
The tree level result for the form factors of the nucleon decay
are available \cite{Aoki:1999tw}. 
The ``direct method'', which uses three point functions 
and various two-point functions, estimates the matrix elements 
for the kinematics of the particular lattice 
simulation. Then they are extrapolated/interpolated to the physical kinematics
with the help of $\chi$PT.

The chiral Lagrangian of the nucleon decay
\cite{Claudson:1982gh} involves only two additional 
low energy parameters to the ordinary three-flavor baryon chiral Lagrangian
at leading order.
Measuring these parameters on the 
lattice and combining with the other parameters of the baryon 
chiral Lagrangian, all the matrix elements of nucleon decay can, in 
principle, be calculated. For the proton to $\pi^0$ decay as an example, 
the relevant form factors read
\begin{eqnarray}
 W_0^{RL} (p\to\pi^0) & = & \alpha (1+D+F)/ \sqrt{2}f,
 \label{eq:chl_p2pi0_R}\\
 W_0^{LL} (p\to\pi^0) & = & \beta (1+D+F)/ \sqrt{2}f,
 \label{eq:chl_p2pi0_L}
\end{eqnarray}
where $f$ is the tree level pion decay constant with a normalization
such that the experimental value is $f_\pi\simeq 131$ MeV. $D$ and $F$ are the couplings of 
baryons to the axial current, where the sum of them gives 
the nucleon axial charge: $D+F=g_A$. $\alpha$ and $\beta$ are
specific to the nucleon decay which can be
calculated at leading order through the proton to  
vacuum matrix element of the operators,
\begin{align}
 \alpha P_L u_p  = 
  \langle 0 | {\mathcal O}^{RL}_{udu} | p\rangle, &&
  \beta P_L u_p = \langle 0 | {\mathcal O}^{LL}_{udu} | p\rangle,
  \label{eqn:alphaL}
\end{align}
\begin{align}
 \alpha P_R u_p =
  -\langle 0 | {\mathcal O}^{LR}_{udu} | p\rangle, &&
  \beta P_R u_p = -\langle 0 | {\mathcal O}^{RR}_{udu} | p\rangle,
  \label{eqn:alphaR}
\end{align}
where Eq.~(\ref{eqn:alphaR}) is obtained from Eq.~(\ref{eqn:alphaL}) by
parity transformation.
We fix the phase definition such that $\alpha$ and $\beta$ are
real and $\alpha<0$. 
As we will later describe, we 
observe $\alpha+\beta\simeq 0$, which is expected because
of the relation,
\begin{equation}
 (\alpha+\beta)\ u_p = -\langle 0|
  \epsilon^{ijk} (u^{T i} C d^j) \gamma_5u^k  |p\rangle,
\end{equation}
which vanishes in the non-relativistic limit and is known to be 
quite small even at small quark masses \cite{Sasaki:2001nf}.

Reduction formulae similar to
Eqs.~(\ref{eq:chl_p2pi0_R}), (\ref{eq:chl_p2pi0_L}) are available for all the
principal matrix elements \cite{Claudson:1982gh,Aoki:1999tw}. These
are summarized in Appendix \ref{sec:app chpt}.
This way of calculating the matrix elements
is referred to as the ``indirect method''.
It has to be noted that the indirect method can have sizable
systematic error, which is difficult to estimate reliably:
\begin{enumerate}
 \item In the $SU(3)_f$ baryon chiral Lagrangian, there are four
       parameters which control the flavor $SU(3)$ breaking effects.
       Two parameters are used to match the baryon masses, and
       enter the nucleon decay matrix elements through baryon masses.
       The other two contribute for the nucleon to pseudo-scalar matrix
       element we wish to extract (and calculate in the direct method),
       but do not contribute to the nucleon to vacuum matrix elements
       required in the indirect method. 
       As we have no means to estimate their values,
       we set these parameters, named $b_i$ $(i=1, 2)$ in
       Refs.~\cite{Claudson:1982gh,Aoki:1999tw}, as $b_i=0$ as is standard
       for such calculations. 
       However, they can naturally be $O(1)$. 
       Setting $|b_i|=1$ for a test,
       the contribution of the breaking term to 
       the relevant form factors is
       estimated as 5--30\% for $N\to K$ decays.
       The worst case is $\langle \eta|(ud)_Ru_L|p\rangle$, where the
       contribution is as large as 200\% \cite{Claudson:1982gh}
       \footnote{As stated, direct calculation of the form factor 
       does not take into account the flavor $SU(3)$ breaking effect
       of $\eta$. Thus, this 200\% correction can be regarded as an
       estimate of the systematic error of direct calculation
       of $\langle \eta|(ud)_Ru_L|p\rangle$.
       The correction for $\langle \eta|(ud)_Lu_L|p\rangle$ is 30\%.}.
       For $N\to \pi$ the effect of these parameters should be negligible.
 \item Even if flavor violation effects were small, i.~e., 
       effects of $b_i$ were negligible, there could be
       appreciable systematic error from the usage of the lowest order
       $\chi$PT.  This is due to the large energy of the pseudoscalar,
       which is at least a half of the nucleon mass: $E_{PS}>m_N/2$
       in the CM frame, while lowest order $\chi$PT is exact only at the zero
       energy (soft pion) limit. 
       Of course, the direct method also relies upon leading order
       chiral perturbation theory at an energy scale around $m_N/2$.
       This is clearly a source of systematic error.
       However, since the extrapolation required is over a much shorter
       distance, it may be expected the systematic error is smaller than
       that for the indirect method.
\end{enumerate} 

The indirect method requires the calculation of only a few two-point
functions on the lattice.
The direct method is superior to the indirect method since for the former
there is no need to assume any parameter in the chiral Lagrangian to be
in any particular range of values. 
Also, as mentioned, the former has less reliance on $\chi$PT.
The coefficients are
determined by fit to the lattice results obtained for each decay process
independently. 
The practical problem of the direct method is that it is 
typically an order of magnitude more demanding in
computation than indirect method for a
similar statistical accuracy.
This is because for the direct method, many types (momenta, masses and
sources) of quark propagators must be solved and a larger temporal lattice
size is needed to accommodate the three point functions.

\subsection{Lattice calculation setup}

We give here brief details 
of the aspects of the lattice calculation which
are common to the following sections. 
More detail will be provided in each section.

We use the domain-wall fermion (DWF)
\cite{Kaplan:1992bt,Shamir:1993zy,Furman:1995ky} action 
for quarks. At the expense of an additional fifth dimension, 
DWF formulation preserves the flavor and chiral 
symmetries of continuum QCD at finite lattice spacing
\cite{Blum:1996jf}.
There are two main reasons to use 
a formulation with good chiral properties:
\begin{enumerate}
 \item Because of chiral symmetry, mixing between operators in different
       chiral multiplets is prohibited. As will be explained
       subsequently, for the particular case of interest this implies
       that the operators are renormalized multiplicatively,
       as in continuum QCD.  Hence, simple and clean handling of the operator
       renormalization is possible.
 \item In lattice gauge simulations, one of the most important sources
       of systematic error is due to finite lattice spacing, $a$.
       Chiral symmetry disallows $O(a)$ scaling violations for both on-
       and off-shell Greens functions, which will participate
       in our estimate of the nucleon decay matrix elements.
       As such, it suggest a mild dependence of any observables on $a$.
       Indeed, DWF has shown good scaling behavior in various hadronic quantities
       \cite{Blum:1997mz,Blum:2000kn,AliKhan:2001wr,Aoki:2002vt,Aoki:2005ga}.
       An important consequence of the fact that the propagator is off-shell
       improved is that non-perturbative renormalization becomes much simpler
       allowing for the possibility of significant reduction
       in systematic error.
\end{enumerate}
For a fifth dimension of finite extent, there is still some
explicit breaking of chiral symmetry by the DWF action. However, it may be
hoped that such breaking is small enough for computationally reasonable
extents of the fifth dimension that it may be either ignored, or treated as
a small correction.
Indeed, as it will be shown later, mixing of the
operators is absent to a high degree of accuracy with 
a finite, affordable size of the fifth dimension.

A convenient quantity to parameterize the size of the explicit chiral 
symmetry breaking is the residual mass, $\mres$, which acts as an additive
renormalization to the fermion mass $m_f$, and is defined 
by considering the  
Ward-Takahashi (WT) identity \cite{Furman:1995ky}, 
\begin{eqnarray}
 \langle {\mathcal A}_\mu^a(x) J_5^b(0) \rangle & = &
  2 m_f \langle J_5^a(x) J_5^b(0) \rangle
  + 2 \langle J_{5q}^a(x) J_5^b(0) \rangle
  + i \langle \delta^a J_5^b(0) \rangle,\\ 
  \mres & = & \frac{\sum_{\vec{x}}\langle J_{5q}^a(\vec{x},t)
  J_5^b(0)\rangle}{\sum_{\vec{x}}\langle J_5^a(\vec{x},t)
  J_5^b(0)\rangle},\; (t\gg 1),
  \label{eqn:mres}
\end{eqnarray}
where ${\mathcal A}_\mu^a$ is the flavor-non-singlet axial current
defined using all bulk 5d fermion field and is a point-split bilinear
operator in 4d sense. 
$J_5$ is the pseudoscalar field constructed from the 4d quarks
which are defined from the 5d domain-wall fermion field by taking
the values at both walls, $s=0$, $L_s-1$ ($L_s$ is the size of the fifth
dimension).
$J_{5q}$ is similar to $J_5$, but, is defined with the fermion field
located at the mid-points ($s=L_s/2-1, L_s/2$) of the fifth dimension. 
The definition of $m_{\rm res}$ is such that the WT identity
takes the same form as that in the continuum, with a shifted mass of
$m_f+m_{\rm res}$
\footnote{For the ratio of correlators given on the right-hand side
of Eq.~(\ref{eqn:mres}) to be represented by a single number, we
are assuming that the effect of $J_{5q}$ on the 4-dimensional 
physics is local, and we are neglecting effects of $O(a^2)$.}.
Hence, ${\mathcal A}_\mu^a$  is often 
called the ``conserved axial current''. Note that, by the WT identity, 
at $m_f=-\mres$ the pion mass must vanish, $m_\pi^2\to 0$ 
as $(m_f+\mres)\to 0$. 
For further details of our conventions and notation see \cite{Blum:2000kn}.

It has been demonstrated in quenched calculations that the chiral
properties of DWF are improved on configurations generated by improved
gauge actions 
\cite{AliKhan:2000iv,Aoki:2002vt}. In particular the DBW2 gauge action
\cite{Takaishi:1996xj,deForcrand:1999bi} is superior in the smallness
of $\mres$ at a given $L_s$ \cite{Aoki:2002vt}.
We use this DBW2 gauge action for both quenched and dynamical 
fermion simulations. 
The drawback of using the DBW2 action is that the sampling of
different topological sectors becomes harder for finer lattices,
which has already been observed for $a\simeq 0.1$ fm \cite{Aoki:2002vt}.
The main quenched calculation in this study uses a coarser lattice, with
$a=0.15$ fm, where the sampling problem is absent. We also use 
an ensemble with $a=0.1$ fm, where we are able to overcome the sampling
problem  by setting larger separation in Monte Carlo time.

Because of the smallness of the explicit chiral symmetry breaking,
${\mathcal A}_\mu^a$ can be treated as a (partially) conserved current
to a good precision. As such, one can calculate 
the renormalization of the naive local (non-conserved) axial vector 
current, $A_\mu^a$, by taking the ratio
\begin{equation}
 Z_A \simeq \frac{\langle \sum_{\vec{x}}{\mathcal A}_\mu^a(\vec{x},t) J_5^a(0) \rangle}
  {\langle \sum_{\vec{x}} A_\mu^a(\vec{x},t) J_5^a(0) \rangle},\; (t\gg 1).
  \label{eq:Z_A}
\end{equation}
Here, due to the point split nature of
${\mathcal A}_\mu^a(x)$, a linear combination of the displaced operators
should be used to get rid of 
$O(a)$ error and to reduce $O(a^2)$ error
\footnote{As with the extraction of the residual mass, for this
equation to be true we are assuming locality and neglecting
$O(a^2)$ effects, and so there is some minimum value of $t$
before which $Z_A$ may not be extracted.}. 
Precise
details of this technique can be found in \cite{Blum:2000kn,Aoki:2002vt}.
The axial current renormalization calculated in this way is used
as a building block of our non-perturbative renormalization 
of the nucleon decay operators.

\begin{table}
 \caption{Domain-wall fermion simulation parameters for
 quenched ($N_f=0$) and unquenched ($N_f=2$) runs
 with DBW2 gauge action.
 Lattice spacing $a$ indicates the approximate value with the $\rho$ mass input
 at the chiral limit. 
 Unrenormalized, approximate value of $\mres$ is presented in physical
 unit. More detailed value of them will be shown in the later sections.
 ``\#configs'' shows the number of 
 configurations analyzed in either matrix element (ME) or
 non-perturbative renormalization (NPR) calculation.
 Unquenched simulation takes three set of degenerate masses.
 }
 \label{tab:params}
 \begin{tabular}{ccccccccc}
  \hline
  \hline
  $N_f$ & $a$ [fm] & $6/g^2$ & $L_\sigma^3\times L_\tau$ & $L_s$ & $M_5$ &
  $\mres$ [MeV] & \#configs(ME) & \#configs(NPR) \\
  \hline
  {\bf 0} & {\bf 0.15} & {\bf 0.87} & {\bf $16^3\times 32$} & {\bf 12} &
  {\bf 1.8} & {\bf 1.3} & {\bf 100} & {\bf 51}\\
  0 & 0.1  & 1.04 & $16^3\times 32$ & 16 & 1.7 & 0.04 & 400 & 55\\
  \hline
  2 & 0.12 & 0.8 & $16^3\times 32$ & 12 & 1.8  & 2.3 & 94$\times$3 & 37--47\\
  \hline
  \hline
 \end{tabular}
\end{table}

Table \ref{tab:params} summarizes the simulation parameters.
Our main effort is devoted to the quenched simulation ($N_f=0$)
at $a=0.15$ fm, where we perform both direct and
indirect measurements of the nucleon decay matrix elements.
Finer lattice with $N_f=0$ will be used to
discuss the finite lattice spacing and volume effects.
An investigation using dynamical fermion for the indirect calculation
of nucleon decay matrix element 
has also been performed to evaluate the quenching effect.

%% file: text_sections/npr.tex
%
%

\ifnum\theOutline=1
\begin{outline}
\item Overview
\item Operator property
\item NPR formulation, kinematics
\item Continuum renormalization and matching
\item lattice parameters and results
\end{outline}
\fi

In order to relate matrix elements obtained with the bare lattice
operator to the continuum counterpart in a given renormalization
scheme, one needs a well prescribed way to renormalize the lattice
operator.  In the literature perturbation theory has been used for
the  renormalization of the nucleon decay operators on the lattice.
However, lattice perturbation theory suffers 
from bad convergence, primarily due to 
the tadpole contribution.
Mean field improved perturbation theory 
\cite{Lepage:1992xa}
works much better.
However, when compared to non-perturbative extractions
there are some operators for which it is inaccurate.
Moreover, the definition is ambiguous.
To be precise:
there are several ways to select the mean 
field factor, and the resulting
renormalization factor sometimes depends on the choice, and it is
indeed the case for the nucleon decay operator for DWF
\cite{Aoki:2002ji}. These problems are naively expected to
disappear in the continuum limit. However, it is preferable to 
avoid such an ambiguity. 

Non-perturbative renormalization (NPR) solves these problems.
In this work we employ the 
non-perturbative, MOM-scheme, renormalization technique of the
Rome-Southampton group  \cite{Martinelli:1995ty}, which has previously been
successfully used in conjunction with DWF 
\cite{Blum:2001sr,Blum:2001xb} in the context of the
renormalization of the flavor non-singlet quark bilinear operators
and four-quark operators relevant to Kaon physics. Since the
MOM-scheme can be applied to any regularization,
it is also referred to as the
regularization independent (RI/MOM)
scheme. Our approach in this work is to
use NPR on the lattice to extract the renormalization 
factors defined at some scale in the MOM-scheme, and then
match the MOM-scheme to the $\overline{MS}$-scheme, which is
more commonly used for the calculation of Wilson coefficients, using
continuum perturbation theory.
The renormalized operators 
are regularization independent 
up to discretization error for the lattice calculation,
e.g.~$O(a)$ for Wilson, $O(a^2)$ for DWF, and up to
the truncation error in continuum perturbation theory.
Since we are relying -- in part -- on continuum perturbation theory,
we must work at a momentum scale for which it is applicable, and so we need
this renormalization scale to be much greater than $\Lambda_{QCD}$.
This may cause a problem because the lattice discretization error grows
as the momentum becomes larger. This is the well-known
window problem: is it possible to have a region of momentum 
such that $\Lambda_{QCD}\ll |p| \ll \pi/a$. 
We will give an estimate of the systematic error due to the window
problem later. 

The anomalous
dimension of the nucleon decay operator has been calculated up to two loops
\cite{Nihei:1995tx} 
using naive dimensional regularization (NDR) in QCD.
We use this result both
in the scheme-matching calculation,
and to factorize the proper scale dependence of the NPR-MOM 
renormalization factor.  
In general, scheme dependence appears at the next-to-leading
oder (NLO), and so  the one-loop matching factor which relates the
MOM scheme to $\msbar$, NDR is needed for the  complete NLO treatment of the
operator renormalization.
The result for this will be presented in Appendix \ref{sec:app pr}.

\subsection{Operator mixing}

Given the good chiral symmetry of DWF, the mixing of the operators with
different chirality is expected to be suppressed. However,
it is instructive to first enumerate the allowed mixings if chiral
symmetry is not assumed.
Since this discussion uses only the rotational, parity, and the vector
flavor symmetry of lattice QCD, it also gives the operator mixing
structure for Wilson fermions.

It is convenient to introduce the operator basis which mimics those
commonly used for the four-Fermi operators in the weak effective Hamiltonian.
All the operators can be written in the form
\begin{equation}
\mathcal{O}^{\Gamma\Gamma'}_{uds}= \epsilon^{ijk}(u^{iT}
C \Gamma d^j) \Gamma' s^k.
\end{equation}
which should be Lorentz spinor, thus all suffixes other than single spin index
must be contracted. 
Here, again $u$, $d$, and $s$ are not necessarily labeling the real flavors.
With a notation:
$S=1$, $P=\gamma_5$,
$V=\gamma_\mu$, $A=\gamma_\mu\gamma_5$,
$T=\sigma_{\mu\nu}=\frac{1}{2}\{\gamma_\mu,\gamma_\nu\}$,
$\tilde{T}=\gamma_5\sigma_{\mu\nu}$, 
we have $\Gamma\Gamma'=$ $SS$, $PP$, $VV$, $AA$, $TT$ for the negative
parity ($\Prt{-}$) operators, and $SP$, $PS$, $VA$, $AV$,
$T\tilde{T}$ for the positive parity ($\Prt{+}$) operators
\footnote{In this notation, the $P\Gamma'$ operator consists 
of scalar diquark.}.
There is another global symmetry which is useful in classifying 
these operators: switching ($\Swt{}$) $u$ and $d$ is a symmetry of 
the Lagrangian if they are degenerate in mass. 
Under a switching transformation, an operator
comes back to itself with possible change of sign depending on $\Gamma$ that
connects the spin indices of $u$ and $d$. Recalling $(C\Gamma)^T=-C\Gamma$ for 
$\Gamma=S$, $P$, $A$ ($\Swt{-}$) and $(C\Gamma)^T=+C\Gamma$ for
$\Gamma=V$, $T$ ($\Swt{+}$), we have four different operator
groups as shown in Table \ref{tab:operator}.
\begin{table}
 \caption{Classification of the nucleon decay three quark operator
$\mathcal{O}^{\Gamma\Gamma'}_{uds}$ by parity ($\Prt{}$) and 
switching  ($\Swt{}$) ($u\leftrightarrow d$).}
 \label{tab:operator}
 \begin{center}
  \begin{tabular}{c|ccc|cc}
   \hline
   \hline
   & & $\Swt{-}$ & & \multicolumn{2}{c}{$\Swt{+}$} \\
   \hline
   \hline
   $\Prt{-}$ & $SS$ & $PP$ & $AA$ & $VV$ & $TT$\\
   $\Prt{+}$ & $SP$ & $PS$ & $-AV$ & $-VA$ & $T\tilde{T}$\\
   \hline
   \hline
  \end{tabular}
 \end{center}
\end{table}
Operators in different blocks do not mix each other.

These operators have the following properties:
\begin{enumerate}
 \item There is a trivial relation between operators with 
       different parity in the same column in Table \ref{tab:operator}:
       ${\mathcal O}(\Prt{+})=\gamma_5{\mathcal O}(\Prt{-})$.
       This means that there is a one to one mapping 
       between the parity negative and parity positive operators
       such that the renormalization matrices are identical.
 \item The five operators in ${\mathcal O}(\Prt{-})$ 
       form a complete set of operators made of
       $u$, $d$ and $s$ with any ordering.
       This follows from the fact that any such operator 
       can be rewritten, by Fierz transformation, as a linear combination of
       the operators $\mathcal{O}^{\Gamma\Gamma'}_{uds}$.
 \item As our target operator is $\Gamma\Gamma'=P_{R/L}P_{L}$ 
(Eqs.~(\ref{eq:operator}), (\ref{eq:me_parity})), 
we may neglect $\Swt{+}$ sector from
possible mixing candidates. Then for each parity (chirality), we need to
consider only three operators for mixing.
 \item  Operators of the type $udu$ are renormalized in the same
       way as $uds$. A simple way to see this is
       to note that the calculation of these renormalization factor using the 
       Rome-Southampton NPR method is identical. This will be shown below. 
\end{enumerate}

From now on, we can concentrate on the $\Swt{-}$ sector.
Our renormalization convention is,
\begin{equation}
 \mathcal{O}^a_{ren} = Z^{ab}_{ND} \mathcal{O}^b_{latt},
\end{equation}
where $a$ and $b$ stand for possible $\Gamma\Gamma'=$ 
$SS$, $PP$, $AA$ for $\Prt{-}$.
$Z^{ab}_{ND}$ is a $3\times 3$ renormalization matrix.
As mentioned above, an identical matrix applies for $\Prt{+}$ operators,
$SP$, $PS$, and $AV$.
The chirality basis, which is more convenient to match the lattice 
operators
with those used in the principal matrix elements,
is made of following three operators,
\begin{eqnarray}
 LL & = & \frac{1}{4}(SS+PP) - \frac{1}{4}(SP+PS),\\
 RL & = & \frac{1}{4}(SS-PP) - \frac{1}{4}(SP-PS),\\
 A(LV) & = & \frac{1}{2}AA - \frac{1}{2}(-AV).
\end{eqnarray}
The renormalization matrix transforms under this
basis change as,
\begin{eqnarray}
 Z^{chiral}_{ND} & = & T\ Z^{parity}_{ND}\ T^{-1},\\
  T & = & \left(\begin{array}{rrc}
	   1/4 & 1/4 & 0\\
	   1/4 & -1/4 & 0\\
	   0 & 0 & 1/2
	  \end{array}\right),
\end{eqnarray}
where $Z^{chiral}_{ND}$ is for the basis operators $LL$, $RL$, $A(LV)$,
while $Z^{parity}_{ND}$ for $SS$, $PP$, $AA$.
If there is no explicit chiral symmetry breaking 
by the action used, $Z^{chiral}_{ND}$ is a diagonal matrix.

Taking the above into consideration, we may contrast the
situation when using Wilson and Domain Wall fermions:
In the Wilson fermion case, 
lattice operator, for example $LL$ operator, is renormalized
at one loop \cite{Bowler:1988us,Aoki:1999tw} as
\begin{equation}
 {\mathcal O}^{\MSbar}_{LL} (\mu) =
  \left(1+\frac{\alpha_s}{4\pi}[4\log(\mu a)+\Delta]\right)
  {\mathcal O}^{latt}_{LL}
  + \frac{\alpha_s}{4\pi} (C_{RL} {\mathcal O}^{latt}_{RL}
  + C_{A(LV)} {\mathcal O}^{latt}_{A(LV)}),
\end{equation}
where, $\Delta$, $C_{RL}$ and $C_{A(LV)}$ are scale independent constants.
The discussion here has shown that these three operators in {\it rhs} are
all that would appear even to all order of $\alpha_s$.
In the DWF case, the mixing between  operators with different chirality
(off-diagonal elements of $Z^{chiral}_{ND}$)  
is highly suppressed. This follows from the discussion 
which makes use of the low energy effective theory of DWF \cite{Blum:2001sr}.  
Applying the same procedure, it can be simply shown 
that the off-diagonal elements are suppressed by a factor of $\mres^2$
($< O(10^{-6})$ for all our parameters). 

\subsection{NPR formulation}

To calculate the renormalization factors using the MOM-scheme 
\cite{Martinelli:1995ty,Blum:2001sr}
we first calculate the Greens function of the particular
operator in question with three external
quark states in Landau gauge,
\begin{equation}
 G^{a}(x_0,x_1,x_2,x_3)=\langle {\mathcal O}_{uds}^{a}(x_0)
 \bar{u}(x_1) \bar{d}(x_2) \bar{s}(x_3) \rangle.
\end{equation}
We set $x_0=0$. A Fourier transformation is then 
performed on the three  external quark legs with the same 
momentum $p$, which are
then amputated to obtain the vertex function, 
\begin{equation}
 \Lambda^{a}(p^2) = \mbox{F.T.}\ G^{a}(0,x_1,x_2,x_3) |_{Amp}.
\end{equation}
Writing the tensor indices explicitly, the renormalization condition of
the MOM scheme reads, 
\begin{equation}
 P^a_{ijk\;\beta\alpha\;\delta\gamma} \cdot
  Z_q^{-3/2} Z^{bc}_{ND} \Lambda^c_{ijk\;\alpha\beta\;\gamma\delta}
  = \delta^{ab},
  \label{eq:condition}
\end{equation}
where $Z_q$ is the quark wave function renormalization; $i, j, k$ are 
color indices, $\alpha$, $\beta$ and $\gamma$, $\delta$ 
are Dirac indices associated with $\Gamma$, $\Gamma'$ respectively. 

The projection matrix $P^a$ is chosen such that 
Eq.~(\ref{eq:condition}) holds for the free field case with $Z_q=1$,
$Z^{ab}_{ND}=\delta^{ab}$. Then, 
\begin{eqnarray}
 P^{SS} & = & \frac{1}{96} \epsilon^{ijk}(C^{-1})^{\beta\alpha}
  \delta^{\delta\gamma}\\
 P^{PP} & = & \frac{1}{96} \epsilon^{ijk}(\gamma_5 C^{-1})^{\beta\alpha}
  \gamma_5^{\delta\gamma}\\
 P^{AA} & = & \frac{1}{384} \epsilon^{ijk}(\gamma_5 \gamma_\mu
  C^{-1})^{\beta\alpha} (\gamma_5 \gamma_\mu)^{\delta\gamma}.
\end{eqnarray}
To simplify notation, we define the matrix $M$ as,
\begin{equation}
 M^{ab} = \Lambda^a_{ijk\;\alpha\beta\;\gamma\delta} \cdot
  P^b_{ijk\;\beta\alpha\;\delta\gamma},
\end{equation}
which is equal to $Z_q^{3/2}(Z^{-1}_{ND})$ up to systematic errors.

The treatment of $Z_q$ needs care \cite{Blum:2001sr},
as its definition naturally involves the derivative with respect to
the momentum, and the lattice momentum cannot be continuous.
Here we exploit the accurate determination of $Z_A$ 
from the hadronic matrix element (Eq.~(\ref{eq:Z_A})).
The bilinear vertex 
function of the local axial current $\Lambda_A$
calculated in the MOM-scheme yields
$Z_q/Z_A$. Thus, the ratio $\Lambda_A^{3/2}/M^{aa}$ will give 
$Z_{ND}/Z_A^{3/2}$. $Z_{ND}$ is calculated with these 
two measurements without directly dealing with $Z_q$.

As $Z_A$ has no scale dependence in the continuum,
it must not have scale dependence except for that brought by
the discretization error ( which starts at $O(p^2a^2)$)
at the finite lattice spacing. The ratio $Z_{ND}/Z_A^{3/2}$,
then, has the same scale dependence 
as the nucleon decay operator up to $O(p^2a^2)$ scaling 
violations.

\subsection{Scheme matching and RG running}

As it is clear in the above discussion, we need to know the scale
dependence of the renormalized nucleon decay operator to separate 
it out from the potential lattice artifacts. Our goal 
is to quote values for the matrix elements of interest in 
the $\msbar$, NDR scheme at some scale, $\mu$,
\begin{equation}
 {\mathcal O}^{\MSbar} (\mu) = U^{\MSbar\leftarrow latt}(\mu)
  {\mathcal O}^{latt},
  \label{eq:op_lat_to_msbar}
\end{equation}
where $U^{\MSbar\leftarrow latt}(\mu)$ is the renormalization factor
needed.
We are using a two-step renormalization procedure: first renormalize
with the MOM-scheme at scale $p$, and then match with the
$\msbar$ scheme at $\mu$. This leads to the equation
\begin{equation}
 U^{\MSbar\leftarrow latt}(\mu) = U^{\MSbar}(\mu;p)
  \frac{Z^{\MSbar} (p)}{Z_{cont}^{MOM}(p)} Z_{latt}^{MOM}(p).
  \label{eq:lat to msbar}
\end{equation}
The $Z_{cont}^{MOM}$ and $Z_{latt}^{MOM}$ are MOM-scheme 
factors calculated using
continuum perturbation theory and NPR on the lattice respectively.
$Z^{\MSbar} (p)$ is the continuum $\msbar$ renormalization factor,
$U^{\MSbar}(\mu;p)$ is the renormalization group  evolution factor from
scale $p$ to $\mu$ in $\msbar$.
$Z^{\MSbar} (p)/Z_{cont}^{MOM}(p)$ is the matching factor from 
$\msbar$ to MOM at scale $p$.

The anomalous dimension of the nucleon decay operator, which enters
$U^{\MSbar}(\mu;p)$, has been calculated up to two loops in $\msbar$,
NDR scheme \cite{Nihei:1995tx}. 
The anomalous dimension reads,
\begin{eqnarray}
 \gamma & = & \gamma_0 \frac{\alpha_s}{4\pi} + \gamma_1
  \left(\frac{\alpha_s}{4\pi}\right)^2,\\
 \label{eq:ND ano dim}
  \gamma_0 & = & -4,\;
  \gamma_1 = -\left(\frac{14}{3}+\frac{4}{9}N_f-4\Delta\right),
\end{eqnarray}
where $\Delta=0$ for $LL$, $-10/3$ for $RL$ operator,
and $N_f$ is the number of active flavors
\footnote{Typo in \cite{Nihei:1995tx} corrected by private
communication with Nihei.}.
The $\msbar$ evolution factor reads
\begin{eqnarray}
  U^{\MSbar}(\mu;p) & = &
  \left[\frac{\alpha_s(\mu)}{\alpha_s(p)}\right]^{\gamma_0/2\beta_0}
  \left[1+\left(\frac{\gamma_1}{2\beta_0}-\frac{\beta_1\gamma_0}{2\beta_0^2}
	  \right)
  \frac{\alpha_s(\mu)-\alpha_s(p)}{4\pi}
  \right],
  \label{eq:msbar running}\\
 \beta_0 & = & 11-\frac{2}{3}N_f,\; \beta_1 = 102-\frac{22}{3}N_f.
\end{eqnarray}

The matching factor is calculated to one loop in continuum perturbation
theory. The MOM-scheme calculation should be performed with the same
kinematics and in the same gauge as 
that used on the lattice. Setting
the momenta for the three external quark
fields to be equal, and setting the mass to zero,
the matching factor is obtained as,
\begin{equation}
 \frac{Z^{\MSbar}}{Z^{MOM}} = 1 + \frac{\alpha_s}{4\pi}
  \left[\frac{433}{180}-\frac{1123}{90}\ln 2
   + \xi \left(\frac{587}{180} - \frac{317}{90}\ln 2\right)\right],
  \label{eq:match}
\end{equation}
where $\xi$ is the gauge parameter and $\xi=0$ (Landau gauge) will be
used. See appendix \ref{sec:app pr} for the derivation.
To match the NPR-MOM scheme with $\msbar$ with this formula,
we need to take the chiral limit of
the massive simulation data. As we will see, this can be done very 
precisely, as 
this mass dependence is extremely mild in the NPR data.

We use the two-loop running coupling $\alpha_s(\mu)$ 
with $\Lambda_{\MSbar}$ obtained by Alpha collaboration
for quenched QCD \cite{Capitani:1998mq},
  $\Lambda_{\MSbar}^{(0)} r_0 = 0.602(48)$,
where $r_0$ is the Sommer parameter defined with the static quark
potential $V(r)$ as $r^2\frac{dV}{dr}=1.65$ \cite{Sommer:1994ce}.
The approximate value is $r_0=0.5$ fm from the potential models.
As we set the scale by using the $\rho$ meson, we use our measurements
of $r_0/a$ and $m_\rho a$ and combine with the experiment 
$m_\rho=0.77$ GeV to get the appropriate $\Lambda_{\MSbar}^{(0)}$.

\subsection{Results of the NPR}
\label{subsec:npr_results}

We present here the results of the NPR of nucleon decay operators
for quenched calculation, on configurations generated with
the DBW2 gauge action at $a=0.15$ fm (see, Table \ref{tab:params}).
The NPR study employs
four quark masses  $m_f=0.025$, $0.04$, $0.055$, $0.07$, where
the largest roughly corresponds to the strange quark mass.

\begin{figure}[h]
 \begin{center}
  \includegraphics[width=10cm]{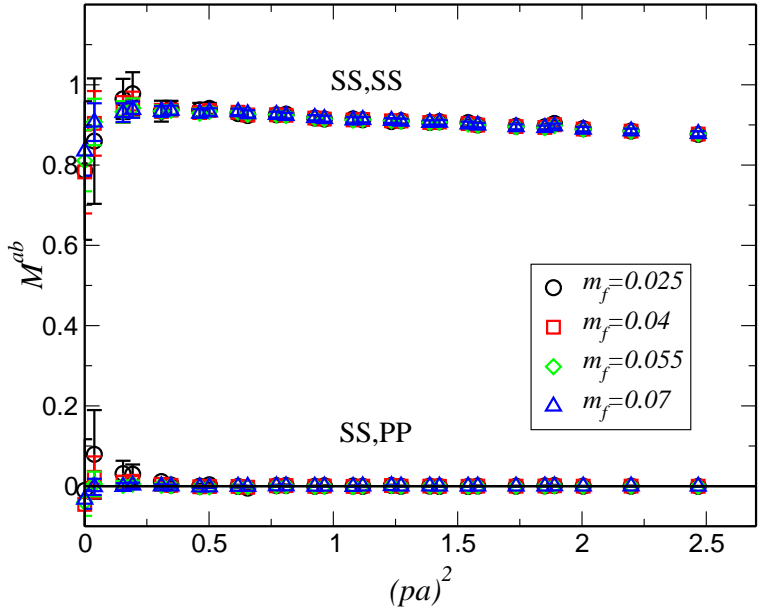}
 \end{center}
 \caption{$SS$ and $PP$ projections of
 the $SS$ operator, $M^{SS,SS}$ and  $M^{SS,PP}$
 as a function of lattice momentum squared for each quark mass.}
 \label{fig:M_SS_PS}
\end{figure}

\begin{figure}[h]
 \begin{center}
  \includegraphics[width=12cm]{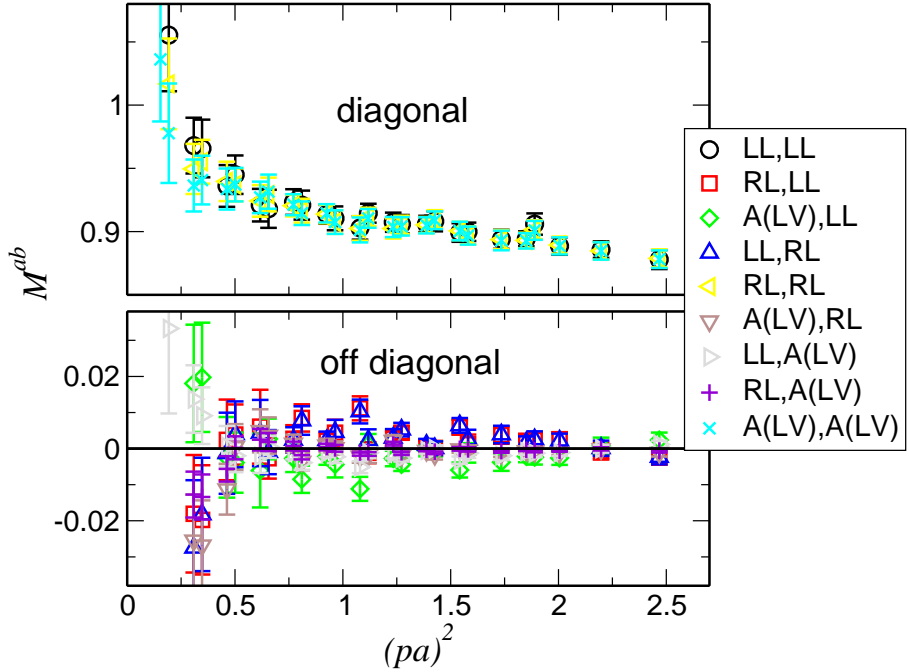}
 \end{center}
 \caption{Mixing matrix $M^{a,b}$ in the chirality basis 
 at the chiral limit $m_f\to -\mres$.}
 \label{fig:Mc_all_0}
\end{figure}

Figure \ref{fig:M_SS_PS} shows the $SS$ and $PP$ projections of
the $SS$ operator, $M^{SS,SS}$ and $M^{SS,PP}$
as a function of lattice momentum squared for all quark masses.
Note that mass dependence is negligible.
Taking the chiral limit ($m_f\to -\mres$) 
using a linear extrapolation
and rearranging to the chirality
basis, one obtains Fig.~\ref{fig:Mc_all_0} for all the elements 
of $M$.
Most of the off-diagonal elements are consistent with zero for 
$(pa)^2>1.2$ within $2\sigma$, while others still remain within
1\% of the diagonal elements, and are thus negligible for 
our extraction \footnote{
It should be noted that we do not ascribe the non-vanishing 
values to be a consequence of explicit chiral symmetry breaking due
to a finite fifth dimension, but a combination of low energy spontaneous
chiral symmetry breaking effects and discretization errors.
}.
As a result,  nucleon decay operator 
$\ndop=\epsilon^{ijk}(u^{iT} CP_{R/L}d^j)P_{L}s^k$ is renormalized
multiplicatively for our domain-wall fermion simulation.

\begin{figure}[h]
 \begin{center}
  \includegraphics[width=10cm]{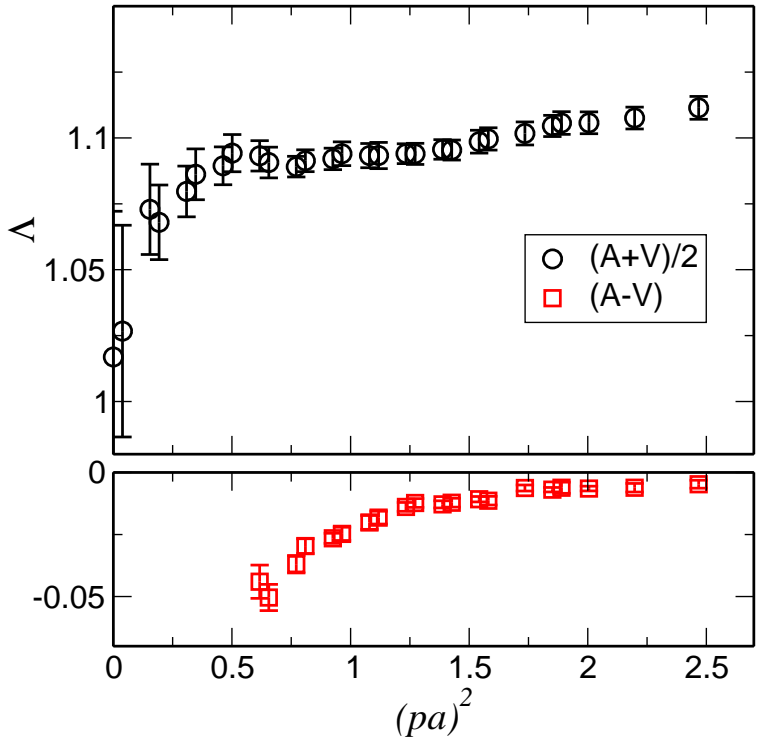}
 \end{center}
  \caption{Average and difference of the axial vector and vector vertex
 functions.}
 \label{fig:AV}
\end{figure}

The next step is to obtain the total 
renormalization factor to relate the lattice operator to the 
$\msbar$, NDR operator, for which we need the value
of $Z_q$. As mentioned previously, we extract this value by
calculating $Z_A/Z_q$ using the Rome-Southampton technique, and
$Z_A$ from hadronic correlators.
For the former, we use the average of vertex 
function of the local axial vector
and vector current operators. The renormalization
constants for these operators are equal in a theory in which chiral
symmetry is only softly broken. This equality should also hold for the
vertex functions at high energies.
At low energies they can differ due to the
spontaneous breaking of chiral symmetry. 
Figure \ref{fig:AV} shows the average and difference of the vertex
function for $A$ and $V$.
The non-zero difference at the value of the momentum which are
accessible to our lattice calculations
indicates that the window where the RI/MOM NPR
can be safely applied is closing for the lattice spacing being used
($a=0.15$ fm). 
This difference may be taken as a measure of the systematic error of the
renormalization constant arising from the closing of the window.
One sees up to 1.5\% effect difference for $p^2\ge 1.2$, which
can be enhanced by the extrapolation, $(pa)^2\to 0$ to 2\%.
We may estimate the systematic error of the nucleon decay
renormalization constants, from this source, as 3\% considering 
the dimension of the operator.
\begin{figure}[h]
 \begin{center}
  \includegraphics[width=10cm]{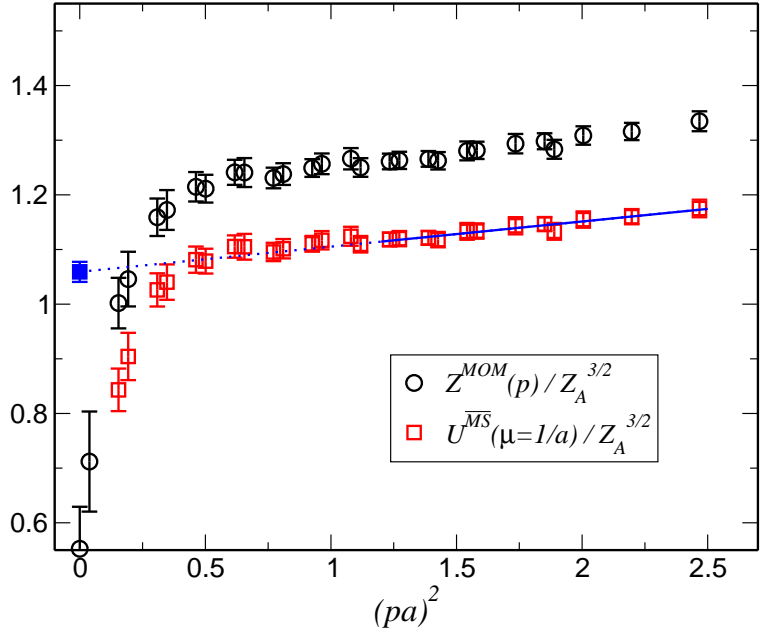}
 \end{center}
 \caption{Renormalization factor of nucleon decay operator 
(${\mathcal O}_{LL}$) normalized by the axial current renormalization
 constant $Z_A$, for a)  (circle) MOM
 scheme $Z^{MOM}_{latt}(p)/Z_A^{3/2}$ as a function of renormalization
 scale $p^2$ and for b) (square) $\msbar$ scheme 
 $U^{\MSbar\leftarrow latt}(\mu=1/a)/Z_A^{3/2}$
 (Eq.~(\ref{eq:op_lat_to_msbar})) as a function of MOM$\to \msbar$ 
  matching scale $p^2$.}
\label{fig:Z_LL}
\end{figure}

Using average of the
axial and vector vertex function,
combined with the vertex function of the nucleon decay operator, the
diagonal elements of $Z_{ND}^{MOM}/Z_A^{3/2}$ 
can be calculated. This is shown in
Fig.~\ref{fig:Z_LL} for ${\mathcal O}^{LL}$ versus
the renormalization scale squared. Using the results of the 
previously discussed matching and running calculations the
$\msbar$ renormalization factor (Eq.~(\ref{eq:lat to msbar}))
at fixed scale $\mu=1/a$, which should, thus be scale invariant, is
shown as squares. This is, again, plotted versus the square of the
scale at which the lattice, MOM-scheme, renormalization calculation 
was performed.
We identify the remaining momentum dependence as 
$O(p^2a^2)$ discretization error. To extract the value at $(pa)^2=0$,
linear function in $(pa)^2$ is used to fit it in the region
$1.2<(pa)^2<2.5$, where the 
non-perturbative effect is expected to be small ($\simle 3\%$) and
higher order effect is negligible.
Combining with the axial current renormalization (Table \ref{tab:qmasses})
obtained with 
Eq.~(\ref{eq:Z_A}), and running from $\mu=1/a$ to $2$ GeV by 
Eq.~(\ref{eq:msbar running}), we get
\begin{equation}
 U^{\MSbar\leftarrow latt}(2\mbox{GeV}) = \left\{
	   \begin{array}{ll}
	    0.751(13)(45) & \mbox{for } {\mathcal O}^{LL},\cr
	    0.755(15)(45) & \mbox{for } {\mathcal O}^{RL},	    
	    \end{array}
       \right.
\end{equation}
where the value for ${\mathcal O}^{RL}$ has similarly been calculated.
The first brackets show statistical error.
The systematic error in the second brackets involves two parts: one is from
the window problem, which we already have estimated as 3\%.
The other is the perturbation theory error arising from truncating the
higher order terms, which are of the order of $\alpha_s^2$.
We take $\alpha_s^2(p)\simeq 0.05$ at the smallest matching momentum
$(pa)^2=1.2$ as the relative systematic error
\footnote{
Another way to estimate the truncation error is to guess from the
difference between LO and NLO. It is about 15\%, which can be seen
in Fig.~\ref{fig:Z_LL}. The difference between NLO and NNLO
is, then, estimated by taking a square as 2\%. This is same order, but,
smaller than 5\% that we have employed.
}. 
Thus, the total systematic error of the NPR is 6\%.

We apply the same procedure to the $a=0.1$ fm DBW2
lattices and get
\begin{equation}
 U^{\MSbar\leftarrow latt}(2\mbox{GeV}) = 
  0.805(9)(32)
\end{equation}
for both ${\mathcal O}^{LL}$ and ${\mathcal O}^{RL}$.
The systematic error from the window problem is negligible.
The perturbation theory error is 4\%, which is counted as the total
systematic error for the renormalization factor. 

%% file: text_sections/me.tex
%
%

\ifnum\theOutline=1
\begin{outline}
\item this section describes the quenched simulation results
\item hadron masses and scale
\item $\alpha$ and $\beta$: $t$-dep, $m_f$-dep, and chiral limit
\item form factors: $t$-dep, $(m_f,q^2)$-dep, physical limit
\item form factors: soft pion limit -> appendix
\item systematic errors
\item comparison $to$ JLQCD
\end{outline}
\fi

In this section the details of the matrix element calculation
in the quenched approximation are given. 

\subsection{Parameters and Lattice Scale}

The matrix elements and related hadron spectrum are calculated on the
quenched DBW2 configurations with lattice spacing $a=0.15$ fm (see Table
\ref{tab:params}). 
The 100 configurations used are
separated by 200 iterations of four overrelaxation 
and one heatbath steps. The quark masses are
$m_f=0.02, 0.04, 0.06, 0.08$.
The strange quark mass point is approximately between the 
largest two masses used.
We use quark propagators with anti-periodic boundary condition in
temporal direction to measure $Z_A$, $m_{res}$.
For all the other quantities we average two quark propagators: one with
periodic, the other with 
anti-periodic boundary condition in temporal direction.
In this way we effectively double the temporal size to $64$,
for which we may safely neglect the effects of the hadrons moving around the
temporal boundary in the three-point function measurements. 

Table \ref{tab:qmasses} shows
the $\pi$, $\rho$, and nucleon masses, $\mres$, local axial current
renormalization $Z_A$, and pion decay constant $f_\pi$ from the
pseudoscalar two point function,
all calculated with the degenerate quarks. 
Non-gaugefixed wall sources \cite{Kuramashi:1994aj} 
are used for $f_\pi$.
The other quantities use a
quark source fixed into Coulomb gauge and  uniformly distributed in a
spatial cubic box, whose size is fixed 
to be $8^3$ to optimize the nucleon signal. 
\begin{table}
 \caption{The residual mass $\mres$, axial current renormalization
 constant $Z_A$, hadron masses from the box source propagator,
 pion decay constant $f_\pi$ from the pseudoscalar two point function
 with gauge unfixed wall source propagator
 for $6/g^2=0.87$ ($a=0.15$ fm) quenched domain-wall fermions.
 The nucleon decay low energy parameters $\alpha$ and $\beta$ with bare
 operator are shown, as well.
 Chiral extrapolation by uncorrelated linear fit in $m_f$ within the
 ``range'' of $m_f$ are performed to get the ``chiral limit'',
 which is defined as $m_f\to-\mres$, except for $m_f\to 0$ to get
the chiral limit of $\mres$ itself.
 For $m_\pi$, squared values are fitted and 
 ``chiral limit'' shows the square root of extrapolated $m_\pi^2$.
 All dimension-full quantities are in lattice units. Errors are by
 jackknife.}
 \label{tab:qmasses}
 \begin{center}
  \begin{tabular}{cllllllll}
   \hline
   \hline
   \multicolumn{1}{c}{$m_f$} & \multicolumn{1}{c}{$\mres (\times 10^{-3})$}
   & \multicolumn{1}{c}{$Z_A$} & \multicolumn{1}{c}{$m_\pi$} & \multicolumn{1}{c}{$m_\rho$}
   & \multicolumn{1}{c}{$m_N$}
   & \multicolumn{1}{c}{$f_\pi$}
   & \multicolumn{1}{c}{$\alpha (\times 10^{-3})$}
   & \multicolumn{1}{c}{$\beta  (\times 10^{-3})$}\\
   \hline
   \hline
   0.02 & 1.171(29) & 0.78405(38) & 0.3060(25) & 0.637(16)  & 0.858(17)  & 0.1107(9) & $-$6.85(66) & 7.56(76)\\
   0.04 & 1.076(25) & 0.78770(28) & 0.4172(22) & 0.6866(69) & 0.9698(83) & 0.1205(9) & $-$7.54(46) & 7.61(44)\\
   0.06 & 1.005(22) & 0.79173(23) & 0.5066(20) & 0.7352(42) & 1.0722(62) & 0.1290(9) & $-$8.28(42) & 8.27(39)\\
   0.08 & 0.955(21) & 0.79609(21) & 0.5849(18) & 0.7829(32) & 1.1651(52) & 0.1368(9) & $-$9.05(42) & 9.02(40)\\
   chiral limit & 1.248(34) & 0.77983(46) & 0.083(11) & 0.587(12) & 0.751(19) & 0.1012(10) & $-$6.03(71) & 6.57(76)\\
   \hline
   range & 0.02$-$0.06 & 0.02$-$0.06 & 0.02$-$0.06 & 0.02$-$0.08 & 0.02$-$0.06 & 0.02$-$0.06 & 0.02$-$0.08 & 0.02$-$0.08\\
   $\chi^2$/dof & 0.14(44)/1 & 0.29(18)/1 & 0.22(15)/1 & 0.01(16)/2 & 0.13(29)/1 & 0.31(19)/1 & 0.005(51)/2 & 0.49(60)/2\\
   \hline
   \hline
  \end{tabular}
 \end{center}
\end{table}
The hadron masses are obtained by the standard two-parameter
correlated fit to the two point functions. $\mres$ and $Z_A$ are 
calculated by the ratio method, Eqs.~(\ref{eqn:mres}), (\ref{eq:Z_A}).
$\mres$ extrapolated to $m_f=0$ is used as the chiral limit point
$m_f\to-\mres$.
In the table, chiral limit values from a linear fit in $m_f$ are also
shown. For pion mass, $m_\pi^2$ is also fit to linear function of $m_f$.
The lower bound of the fitting range is always the smallest mass,
$m_f=0.02$.  For all measurements except for the $\rho$ meson mass we
exclude largest mass $m_f=0.08$ from the chiral fit to
stay in the region of good
linearity. Taking the experimental $\rho$ mass input (0.77 GeV)
and comparing to the chiral limit the lattice cut off is determined,
\begin{equation}
 a^{-1} = 1.312(27)\mbox{ GeV}.
\end{equation}
If we took $f_\pi=0.131$ GeV as an input, we would obtain $a^{-1} = 1.294(13)$
GeV which is consistent with the $\rho$ input.
For the nucleon mass $m_N=0.938$ GeV as an input, $a^{-1} = 1.249(32)$
GeV. 
This scale ambiguity (5\%) between $\rho$ and nucleon input,
likely predominated by combination of extrapolation and quenching
errors, will be used as the systematic error of the matrix elements in
physical units.
Using $a=0.150(3)$ fm from $\rho$ input, the spatial 
lattice size is obtained as $L_\sigma=2.4$ fm, which 
should be large enough for
nucleon with our mass range \cite{Sasaki:2003jh}, and can accommodate
momentum small enough 
to directly reach the physical kinematics region of the nucleon
to pion decay.

The simple linear extrapolation of the pion mass to
the chiral limit gives a non-zero value. Rather than evidence
of numerically significant chiral symmetry breaking which is
not already taken into account by the residual mass, we take this
as a sign of non-analyticity at small mass. 
In fact, fitting using 
a formula including quenched chiral 
logarithms suggested by the quenched
chiral perturbation theory (Q$\chi$PT) leads to a fit
 with reasonable $\chi^2$ under constraint that $m_\pi$ 
vanishes at $m_f=-\mres$.
We use $m_f=-\mres$ as the chiral limit of DWF, thus as the physical
$u$, $d$ masses in our approximation, throughout the paper.
Apart from the pion in the chiral limit, we use leading linear
dependence on the quark mass to interpolate or extrapolate
to the physical points for all the quantities.

The strange mass point $m_f^{(s)}$
is obtained as  $m_\pi^2(m_f^{(s)},m_f^{(s)})=2m_K^2$ 
using the linear fit results to interpolate, 
where $m_K=0.497$ GeV,
\begin{equation}
 m_f^{(s)} = 0.0675(30).
\label{eq:m_s}
\end{equation}
This will be used to interpolate the mass of $\bar{s}$ in the kaon for
the form factors. 

\subsection{Hadronic matrix elements}

The low energy parameters $\alpha$ and $\beta$ 
(Eqs.~(\ref{eq:chl_p2pi0_R}), (\ref{eq:chl_p2pi0_L})) of the nucleon decay
chiral Lagrangian are calculated through the ratio of the two point
functions,
\begin{equation}
 R_{\alpha/\beta}(t)=\frac{\sum_{\vec{x}}\langle \ndop_{udu}(\vec{x},t)\ 
  \overline{J}_{p}(t_0)\rangle} 
  {\sum_{\vec{x}}\langle J_{p}(\vec{x},t)\ \overline{J}_{p}(t_0)\rangle} \sqrt{Z_p}.
  \label{eq:ratio2pt}
\end{equation}
A quark propagator with a cubic box source of $8^3$ volume at $t=t_0$
is used for both-two point functions in
numerator and denominator. The factor
$\sqrt{Z_p}$ is the overlap of 
the proton interpolating field to the normalized proton state,
\begin{equation}
 \langle 0 | J_{p}(\vec{0},0) | p \rangle = \sqrt{Z_p} u_p,
\end{equation}
for the local proton interpolating field $J_p(x)$ and the
proton spinor with the standard relativistic normalization
(see appendix \ref{sec:app notation}, Eq.~(\ref{eq:spin_normalization})).
$Z_p$ could be measured via the amplitude of the single exponential fit to
the point-point proton two point function. 
However, this fit is often problematic. Instead we 
estimate it through the ratio of the proton two-point functions
with point and box source propagators, both with point sink. Asymptotic
value of the ratio gives the ratio of the amplitudes with the point-pion
and box-point propagators.
Given the amplitude of the box-point propagator, which is more accurate
than that with the point-point, $Z_p$ is finally obtained.

\begin{figure}[h]
 \begin{center}
  \includegraphics[width=10cm]{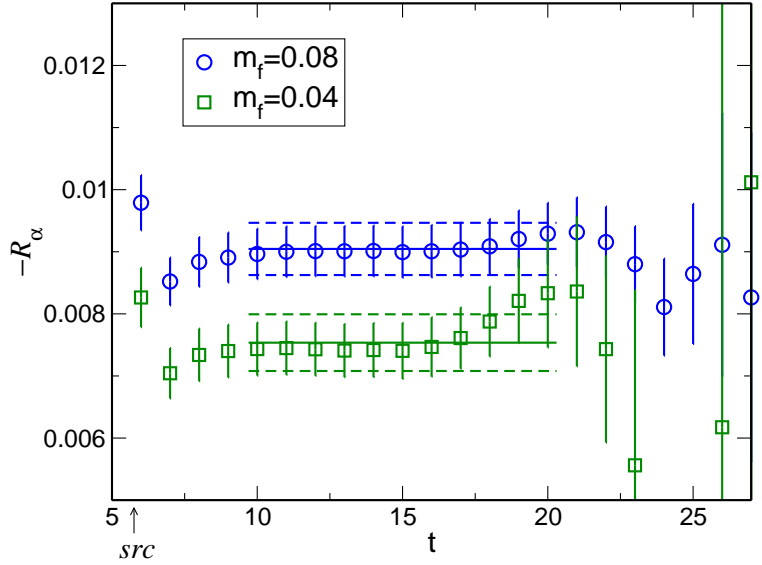}
 \end{center}
 \caption{Ratio $R_{\alpha}(t)$ (Eq.~(\ref{eq:ratio2pt})) for the low
 energy parameter $\alpha$ at $m_f=0.04$ and  $0.08$ on quenched DBW2
 configurations, shown against the operator position ($t$),
 where the proton source is located at $t_0=6$.}
 \label{fig:ratio2pt}
\end{figure}
\begin{figure}[h]
 \begin{center}
  \includegraphics[width=10cm]{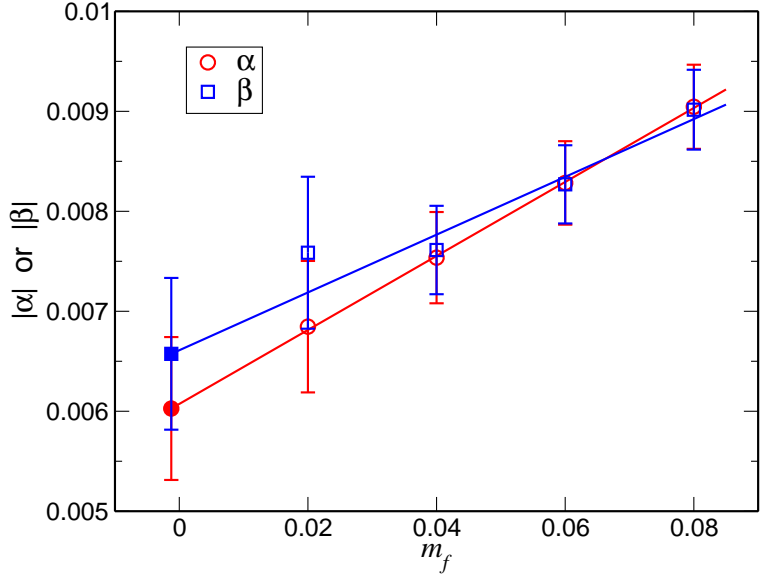}
 \end{center}
 \caption{$|\alpha|$ and $|\beta|$ obtained from the ratio
 $R_{\alpha/\beta}(t)$ as functions of $m_f$ on quenched DBW2 configurations.
 Linear extrapolation is performed to get the values in the chiral
 limit ($m_f\to\mres$). Values are in lattice units and unrenormalized.}
 \label{fig:alpha-mf}
\end{figure}
Figure \ref{fig:ratio2pt} shows the ratio $R_\alpha(t)$ 
(Eq.~(\ref{eq:ratio2pt})) for $m_f=0.04$ and $0.08$. By fitting 
to a constant, $\alpha(m_f)$ is obtained. Figure \ref{fig:alpha-mf} shows 
the fitted $\alpha$ at finite $m_f$, which is extrapolated with 
linear function in $m_f$ to the chiral limit \footnote{
One can take the ratio $\alpha/f_\pi$ ($\beta/f_\pi$) at 
finite mass with using measured $f_\pi$ and extrapolate it
to the chiral limit instead of extrapolating $\alpha$ ($\beta$)
itself. This method has an advantage of making the mass dependence
milder. With our measurement, however, it makes only up to 2\%
difference in the extrapolated values and up to 10\% difference in
the statistical error. 
Apart from the way it is extrapolated, this ratio in the chiral
limit might be more useful 
than the original, since $\alpha$ or $\beta$ enters the form factors 
(Eqs.~(\ref{eq:chl_p2pi0_R}), (\ref{eq:chl_p2pi0_L}), (\ref{eqn:indirect pi0 RL})--(\ref{eqn:indirect eta LL}))
always through the ratio with $f_\pi$.
Also, since the ratio has one dimension less than
the original, the scaling violation is expected to be smaller.
Thus, taking the continuum limit with this ratio has 
an advantage in the discretization where the scaling violation is larger
than the present calculation.
}.
Our estimate of the low energy parameters 
with the operators renormalized at $\mu=2$ GeV is
\begin{eqnarray}
 -\alpha & = & 0.0100(12)(14)(6) \mbox{ GeV}^3,\label{eq:qalpha}\\
 \beta & = & 0.0108(13)(15)(7) \mbox{ GeV}^3,\label{eq:qbeta}
\end{eqnarray}
where the first bracket shows the statistical error from the 
the bare matrix element, second is systematic error from the
scale ambiguity, and the 
third comes from the total error of the renormalization factor.

\begin{figure}[h]
 \begin{center}
  \includegraphics[width=7cm]{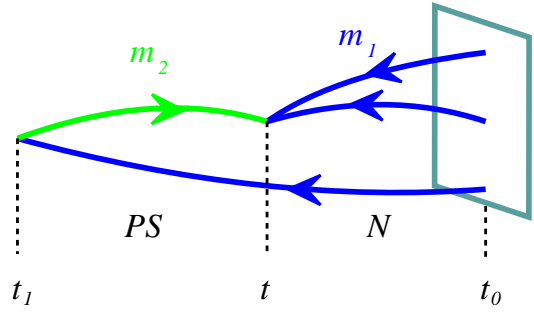}
 \end{center}
 \caption{Diagram of the quark line contraction of the 
 nucleon decay ($N\to PS$) three-point function.}
 \label{fig:3pt_diagram}
\end{figure}

The direct method \cite{Aoki:1999tw} amounts to 
calculating a ratio of the three- and two-point functions. 
For the proton to $\pi$ case, the ratio is 
\begin{equation}
 R_3^{\vec{p}}(t)=\frac{\sum_{\vec{x_1},\vec{x}}e^{i\vec{p}\cdot(\vec{x_1}-\vec{x})}
  \langle J_{\pi}(\vec{x_1},t_1)\ \ndop (\vec{x},t)\ 
  \overline{J}_{p}(t_0)\rangle}
  {\sum_{\vec{x_1},\vec{x}}e^{i\vec{p}\cdot(\vec{x_1}-\vec{x})}
  \langle J_{\pi}(\vec{x_1},t_1) J_{\pi}^{\dagger}(\vec{x},t)\rangle 
  \cdot \sum_{\vec{x}} \langle J_{p}(\vec{x},t)
  \overline{J}_{p}(t_0)\rangle} \sqrt{Z_\pi Z_p}L_\sigma^3, 
  \label{eq:ratio3pt}
\end{equation}
with $Z_\pi$ being the overlap of the pion interpolating field to the
normalized pion state, 
\begin{equation}
 \langle \pi | J_{\pi}^\dagger(0) | 0 \rangle = \sqrt{Z_\pi}.
\end{equation}
Again, the proton interpolating field at $t=t_0$ is made of 
quark fields distributed uniformly in a $8^3$ box.
The associated quark propagators are solved with $m_f=m_{1}$.
In the three-point function, as depicted in Fig.~\ref{fig:3pt_diagram},
two quarks from the nucleon source are
annihilated by the operator at $t=t$. The 
other is a spectator, which is annihilated by the pion interpolating
field at $t=t_1$. To interpolate the pion with momentum, a quark
propagator is solved sequentially with $m_f=m_{2}$ and with the source
equated with the spectator quark propagator at $t=t_1$ with the momentum
projection $e^{i\vec{p}\cdot\vec{x}}$, which acts as an injection 
of the momentum $\vec{p}$. The resulting sequential quark 
propagator is finally contracted as the third quark of the operator at
$t=t$. The momentum $-\vec{p}$ is injected to the operator.

The pion two point function in the denominator is made
with the two quark propagators constructed using non-gaugefixed wall 
sources at $t=t_1$
and with mass $m_{1}$ or $m_{2}$.
One source is always a zero momentum wall, while
the other is made with distribution of $e^{i\vec{p}\cdot \vec{x}}$,
where the momentum $\vec{p}$ should be matched to 
the momentum injected to the sequential quark propagator in the numerator.
Combining the two propagators of this type, the pseudoscalar
operator at the $t_1$ timeslice becomes a local operator 
after averaging over the gauge configurations, as the non-local terms vanish
\cite{Kuramashi:1994aj} by Elitzur's theorem \cite{Elitzur:1975im}.
The non-gaugefixed 
source works well for the pseudoscalar operator and
is superior to the local source in signal/noise ratio.
At time $t$, the quark and the anti-quark are annihilated by
the local pion field $J_\pi^\dagger(x)$ with momentum injection.

We only work with masses satisfying $m_{1}\le m_{2}$, 
which is enough to extrapolate/interpolate to the pion ($(m_{1},
m_{2})\to(-\mres,-\mres)$) or kaon ($-\mres,m_f^{(s)})$) physical point. 
For extrapolation to physical kinematics, $\vec{p}L_\sigma/2\pi=(1,0,0)$,
and $(1,1,0)$ are used, where $L_\sigma=16$ is the spatial size.
Zero momentum $\vec{p}=(0,0,0)$ data
is taken only for $m_{1}=m_{2}$ points to discuss the soft pion
limit in Appendix \ref{sec:app chpt}.

The matrix element is extracted by fitting the plateau of the
ratio as a function of the position of the operator $t$.
Some care is needed to extract the relevant form factor $W_0$
\cite{Aoki:1999tw}. 
The nucleon interpolating field also interpolates the parity
partner of nucleon, which have to be eliminated by the parity projector.
Then naively taking the trace of the ratio, we obtain
\begin{equation}
 Tr\left\{ P_L [W_0 - i \qsl W_q] \left(\frac{1+\gamma_4}{2}\right)
   \right\} = W_0 - i q_4 W_q.
\end{equation}
The irrelevant form factor is calculated as
\begin{equation}
 Tr\left\{ P_L [W_0 - i \qsl W_q] \left(\frac{1+\gamma_4}{2}\right)
   i\gamma_j \right\} = q_j W_q,
\end{equation}
where $q_j=-p_j$, the injected spatial momentum with negative sign.
Combined with the calculated energy transfer $i q_4=E_p-E_\pi$, 
$W_0$ is finally disentangled.

%
\begin{figure}[h]
 \begin{center}
  \includegraphics*[width=10cm]{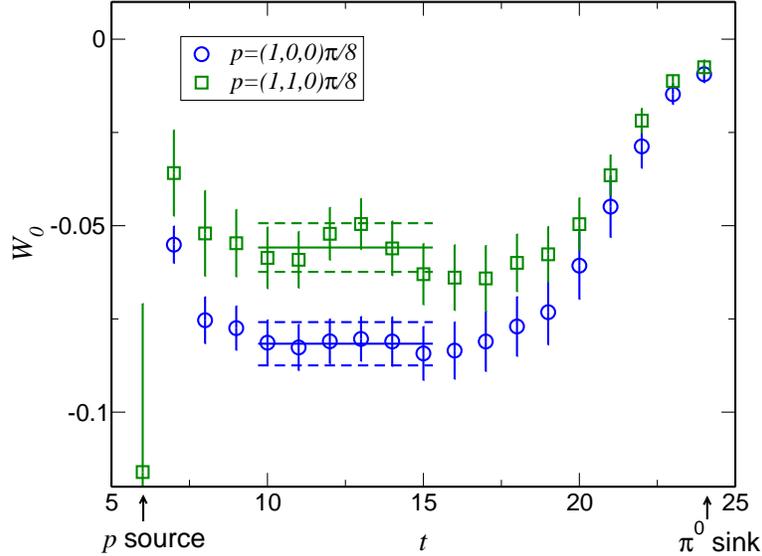}
 \end{center}
 \caption{Ratio Eq.~(\ref{eq:ratio3pt}) for the relevant form factor $W_0$
 of $\langle \pi^0| (ud)_R u_L | p\rangle$ with bare operator and
 in the lattice unit at $m_f=0.06$.}
 \label{fig:ratio3pt}
\end{figure}
\begin{figure}[h]
 \begin{center}
  \includegraphics[width=10cm]{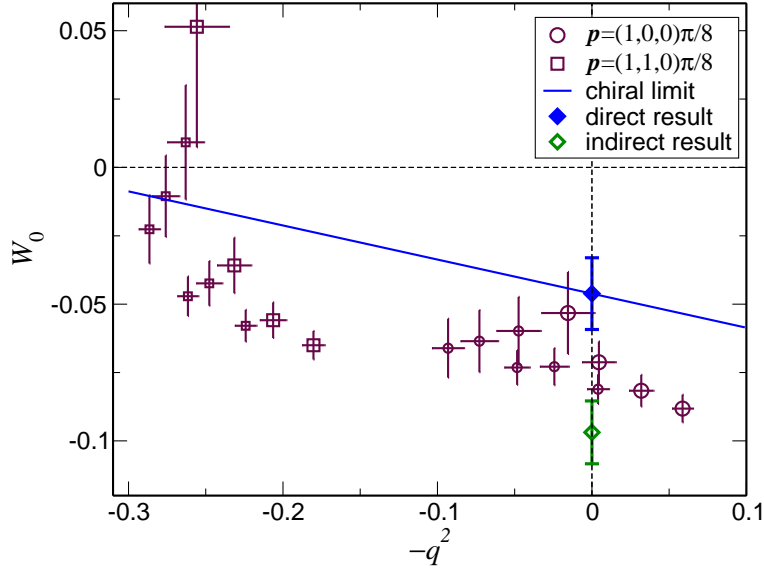}
 \end{center}
 \caption{Relevant form factor $W_0$ of 
 $\langle \pi^0| (ud)_R u_L | p\rangle$ as a function of $-q^2$.
 Circles and squares are the measured data with pion having
 the smallest and the second smallest momentum on the periodic 
 lattice respectively. Larger symbols show the degenerate mass
 points ($m_{1}=m_{2}$), while the smaller indicate non-degenerate
 ($m_{1}\ne m_{2}$).
 The line shows the chiral limit obtained from the fit 
 (Eq.~(\ref{eq:W_0 fit formula})).
 The solid diamond indicates physical kinematics point $q^2\to 0$.
 The open diamond is from the indirect method.}
 \label{fig:W0_RL-qsq}
\end{figure}
\begin{figure}[h]
 \begin{center}
  \includegraphics[width=10cm]{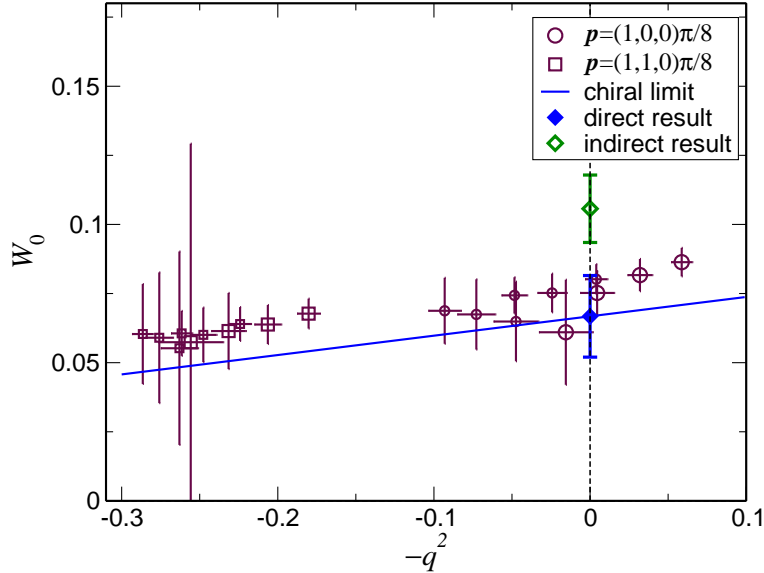}
 \end{center}
 \caption{Same as Fig.~\ref{fig:W0_RL-qsq}, but for 
 $\langle \pi^0| (ud)_L u_L | p\rangle$.}
 \label{fig:W0_LL-qsq}
\end{figure}
Figure \ref{fig:ratio3pt} plots the $W_0$-part of the ratio
Eq.~(\ref{eq:ratio3pt}) as a function of $t$ for
$\langle \pi^0| (ud)_R u_L| p\rangle$ at $m_f=0.06$.
Fitting plateau in the range $10\le t\le 15$, which appears
to be a reasonably good fitting range for all the 
processes  and parameters, 
$W_0(m_f,q^2)$  is obtained and plotted in
Fig.~\ref{fig:W0_RL-qsq} as a function of $-q^2$.
The results are given in lattice units and 
for the bare operator. 
Solid diamond shows the extrapolation to the physical point
($m_1\to -\mres, m_2\to -\mres, q^2\to 0$),
 using  \footnote{
JLQCD \cite{Aoki:1999tw} included a $q^4$ term. 
With our data the effect of including such a term is to
shift $c_0$ within one third of the statistical error.
While $\chi^2/dof$ of the uncorrelated fit increases slightly
by eliminating $q^4$ term, the relative error of $\chi^2$ estimated by
 jackknife  stays in a similar value.}
\begin{equation}
 W_0=c_0 + c_1 (m_{1}+\mres) + c_2 (m_{2}+\mres) + c_3 q^2.
 \label{eq:W_0 fit formula}
\end{equation}
This formula is obtained using leading order $\chi$PT,
expanded in $q^2$ and quark mass $m_q=m_f+\mres$. Quark mass is naturally
interpreted as $m_{PS}^2$, which has similar size as $q^2$ at
the simulated points. The open diamonds are from the indirect method,
Eq.~(\ref{eq:chl_p2pi0_R}) 
taking $f_\pi=0.131$ GeV from experiment. Inputting $f_\pi$ from
our lattice measurement gives a consistent result.
Since the operators are renormalized multiplicatively 
(see Sec.~\ref{sec:NPR}) the results will be unchanged  
after renormalization up to an overall factor.
The indirect calculation estimates $W_0$ to be 
larger by about a factor of 2 than the direct method.
Similar disagreement is seen for the $LL$ operator case, which is shown
in  Fig.~\ref{fig:W0_LL-qsq}.
This difference, though large, is not surprising. At the 
physical kinematics point, where the pion momentum is large
$|\vec{p}|=m_N/2\simeq m_K$, one might expect 
LO $\chi$PT to lose its effectiveness.
On the other hand, the soft pion limit should be described by LO
$\chi$PT exactly. The discussion on this point is given in Appendix
\ref{sec:app chpt}. 

\begin{figure}[h]
 \begin{center}
  \includegraphics*[width=10cm]{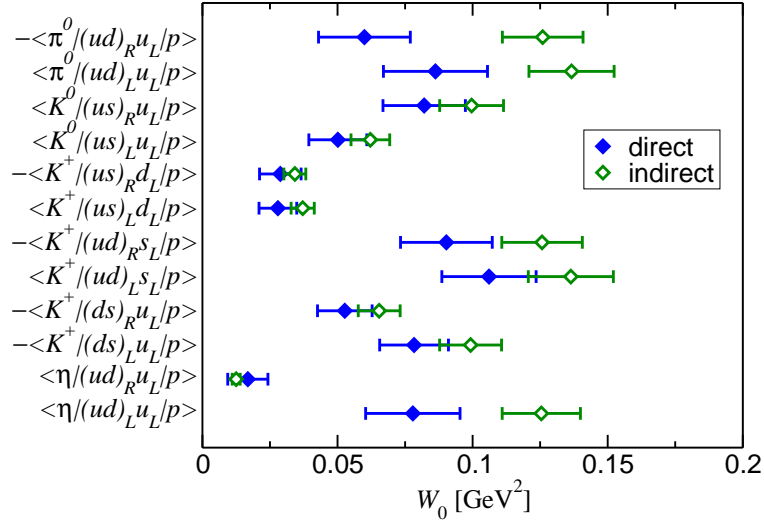}
 \end{center}
 \caption{Relevant form factors of all the principal matrix elements for
 nucleon decay with the quenched   approximation.  Operators are
 renormalized by $\msbar$, NDR scheme at $\mu=2$ GeV. The errors are
 statistical only.}
 \label{fig:W0}
\end{figure}

The same analysis has been carried out for the other 
principal matrix
elements.  All the results for $W_0(m_1,m_2,q^2)$ are summarized
in Table ~\ref{tab:all_W0_1}. The fit results with 
Eq.~(\ref{eq:W_0 fit formula})
are listed in Table ~\ref{tab:all_fit_W0}.
Figure \ref{fig:W0} plots the final results of $W_0$ at the physical
kinematics in physical units for the renormalized operator.
The indirect method estimates the form factor larger than the direct method
for most of the cases. Note that, for several processes,
the difference of the two estimates are significant.

Table \ref{tab:W0} shows the relevant form factors for all the possible
matrix elements. This is the main result of this paper.
We note that the total error
is dominated by the statistical error of the matrix element.

\begin{table}
 \caption{The relevant form factors $W_0$ [GeV$^2$] for all the possible matrix
 elements (Eqs.~(\ref{eqn:p2pi0})--(\ref{eqn:p2eta})), with the operator
 renormalized at $\mu=2$ GeV with $\msbar$, 
 NDR. The first error is total error, which is obtained by adding
 quadrature the statistical error of bare $W_0$ (quoted second error),
 systematic error from the scale, and the total error
 of the renormalization factor. Note again that in the present
 calculation the flavor $SU(3)$ breaking effect of $\eta$ is not
 taken into account.}
 \label{tab:W0}
\begin{tabular}{cc|rcc|rcc}
 \hline
 \hline
 && \multicolumn{3}{c|}{$RL$ or $LR$ operator} &
 \multicolumn{3}{c}{$LL$ or $RR$ operator}\\
 \multicolumn{2}{c|}{\raisebox{2ex}[0pt]{Matrix element}}
 & \multicolumn{1}{c}{$W_0$} [GeV$^2$] & total error & stat.~error 
 & \multicolumn{1}{c}{$W_0$} [GeV$^2$] & total error & stat.~error \\
 \hline
 \hline
 $\langle\pi^0|(ud)u|p\rangle$, & $\langle\pi^0|(du)d|n\rangle$ & $-$0.060 & 0.018 & 0.017 & 0.086 & 0.022 & 0.019\\
 $\langle\pi^+|(ud)d|p\rangle$, & $-\langle\pi^+|(du)u|n\rangle$ & $-$0.085 & 0.026 & 0.024 & 0.122 & 0.030 & 0.027\\
 $\langle K^0|(us)u|p\rangle$, & $-\langle K^+|(ds)d|n\rangle$ & 0.082 & 0.018 & 0.015 & 0.050 & 0.012 & 0.011\\
 $\langle K^+|(us)d|p\rangle$, & $-\langle K^0|(ds)u|n\rangle$ & $-$0.029 & 0.008 & 0.008 & 0.028 & 0.008 & 0.007\\
 $\langle K^+|(ud)s|p\rangle$, & $-\langle K^0|(du)s|n\rangle$ & $-$0.090 & 0.020 & 0.017 & 0.106 & 0.021 & 0.017\\
 $\langle K^+|(ds)u|p\rangle$, & $-\langle K^0|(us)d|n\rangle$ & $-$0.053 & 0.012 & 0.010 & $-$0.078 & 0.015 & 0.013\\
 $\langle\eta|(ud)u|p\rangle$, & $-\langle\eta|(du)d|n\rangle$ & 0.017 & 0.008 & 0.007 & 0.078 & 0.020 & 0.017\\
\hline\hline
\end{tabular}
\end{table}

The values of individual matrix elements are different from those 
obtained for $a^{-1}\simeq 2.3$ GeV with Wilson fermion 
\cite{Aoki:1999tw},  
while ratios of the matrix elements are similar.
We expect our DWF results to be closer to the continuum limit,
as discussed below.

\subsection{Systematic errors}

The major sources of systematic error for this calculation are:
\begin{enumerate}
 \item finite lattice spacing $a$,\label{enum:sys_a}
 \item finite system volume,\label{enum:sys_V}
 \item chiral extrapolation,\label{enum:sys_c}
 \item quenching effects.\label{enum:sys_q}
\end{enumerate}
Among these, \ref{enum:sys_a} and \ref{enum:sys_V} are relatively easier
to address, and, as we see below, there are reasons to believe they 
are small.

To evaluate the systematic errors \ref{enum:sys_a} and \ref{enum:sys_V},
we compare quantities already
shown with those obtained on the finer lattice spacing
and smaller volume. 
We have used a $6/g^2=0.87$ lattice, where lattice spacing
is $a=0.15$ fm, $16^3$ lattice volume corresponds to $(2.4\mbox{ fm})^3$
box.  The finer lattice parameter is a $6/g^2=1.04$, $a=0.1$ fm,
$(1.6\mbox{ fm})^3$ box.

\begin{figure}[h]
 \begin{center}
  \includegraphics[width=8cm]{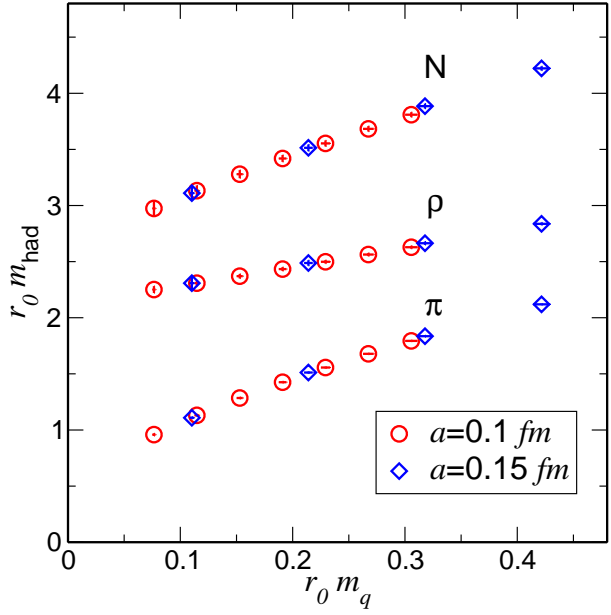}
 \end{center}
 \caption{Hadron masses as a function of quark mass renormalized
 at $\mu=2$ GeV with $\msbar$, NDR. Both axes are normalized by $r_0$.
 Both lattices have $16^3\times 32$ volume. Larger lattice spacing
 $a=0.15$ fm has been used for the nucleon decay matrix element.
 Volume and lattice spacing effects are negligible.}
 \label{fig:qhmasses scaling}
\end{figure}

First, we compare the hadron masses in Table \ref{tab:qmasses} with
those obtained \cite{Aoki:2002vt} on the finer lattice.
Figure \ref{fig:qhmasses scaling} shows the hadron masses normalized
by $r_0$ \cite{Hashimoto:2004rs}
against the quark mass $r_0 Z_m (m_f+\mres)$ renormalized at $\mu=2$ GeV
in $\msbar$, NDR. The quark mass renormalization factors are taken
from ref.~\cite{Dawson:2002nr}, where only $L_s=16$ case is studied.
The difference of $Z_m$ between $L_s=12$ which has been
used in this study and $L_s=16$ should be negligible.
The error shown in the figure is statistical only.%

At larger masses hadrons
are compact, hence the volume effect should be small. The consistency of the
hadron masses at larger quark mass suggests the finite lattice
spacing error is negligible within our accuracy. Furthermore, 
the consistency
is seen all the way down to the smallest mass of $a=0.15$ fm.
This suggests the volume effect is also negligible in the mass range studied.

\begin{figure}[h]
 \begin{center}
  \includegraphics[width=10cm]{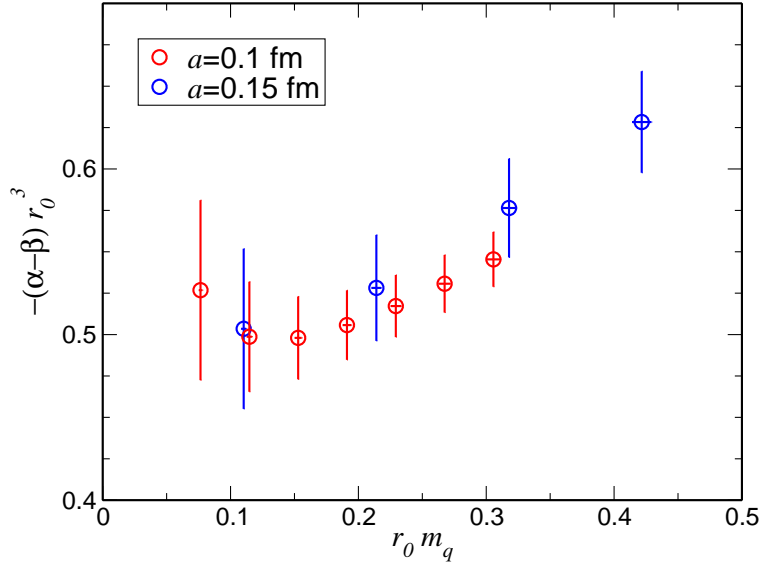}
 \end{center}
 \caption{Difference of the nucleon-decay low-energy parameter
 $-(\alpha-\beta)$ (taking the sign of $\alpha$ and $\beta$ from the
 previously descibed extraction) as a function of quark mass.
 Normalization is same as Fig.~\ref{fig:qhmasses scaling}.
 Consistent results are obtained for the different lattice spacings and
 different spatial volumes.}
 \label{fig:a-b scaling}
\end{figure}
More direct evidence that 
the systematic errors \ref{enum:sys_a} and \ref{enum:sys_V}
 are small
is seen in Fig.~\ref{fig:a-b scaling}, where the difference of
the two low energy parameters $|\alpha-\beta|$ 
is shown in the same way as Fig.~\ref{fig:qhmasses scaling}.
We have used the relation of the proton interpolating field (Eq.~(\ref{eqn:Jp}))
to a nucleon decay operator,
\begin{equation}
 J_p = ( {\mathcal O}^{RL}_{udu} - {\mathcal O}^{LL}_{udu} )
  + ( {\mathcal O}^{RR}_{udu} - {\mathcal O}^{LR}_{udu} ),
\end{equation}
hence,
\begin{equation}
 \langle 0|J_p|p\rangle = (\alpha-\beta) u_p.
\end{equation}
Thus, $\alpha-\beta$ can be 
extracted from the ordinary nucleon two point function 
$\langle J_{p}(x) \overline{J}_{p}(0)\rangle$ as well.
Since the renormalization factors of the nucleon decay operators
of different chirality are same within the error, the average of them
may be used to renormalize $J_p$ to get $\alpha-\beta$.
For the finer lattice results we have reused the $a=0.1$ fm data for
the spectrum study \cite{Aoki:2002vt}. 
The same arguments as the hadron masses lead to the observation that
the systematic errors from lattice spacing and the volume are negligible
also for $\alpha-\beta$.
From this, we can expect that those errors on individual $\alpha$,
$\beta$ and further the relevant form factor $W_0$ are 
negligible compared to the other errors.

Systematic error from the chiral extrapolation may be caused by
the linear extrapolation in quark mass, while in general,
contribution of higher order terms, and in the  special case
of quenching (by Q$\chi$PT), lower order terms, may not be negligible.
Our linear extrapolations for matrix elements and hadron masses (for extracting 
the lattice scale)
have been performed for the mass region where the resulting $\chi^2$ is small.
This does not necessarily mean the leading linear dependence
is sufficient. This can be the case when a term with lower power occurs at 
higher loops in the chiral expansion. An
example of this is the $\rho$ \cite{Booth:1997hk} and nucleon
\cite{Labrenz:1996jy} masses, for which $m_q^{1/2}$ terms appear 
at one loop. The correct quenched results
may only be found when using a proper Q$\chi$PT
formula with data sufficiently good in quality and in quantity to determine
the fitting parameters.  Only after that we can deal with the quench
error by examining the difference to the experiment or by comparing to the
unquenched results. Lack of the Q$\chi$PT knowledge of the matrix elements
as well as the limited quality and quantity for our data
make it difficult to follow this scheme. 

We observed a scale inconsistency between the $\rho$ and nucleon mass
inputs when using the linear chiral extrapolation. This difference
has been taken as 
the systematic error to the matrix elements due to the scale ambiguity.
Since the finite volume and finite lattice spacing errors are
negligible, we expect this scale error is dominated by the chiral
extrapolation and quenching error. 
Whether this estimate is plausible or not 
will be examined by comparing with the results with dynamical fermion
simulation, which will be discussed in the next section.
Note, however, the comparison is done only at a finite lattice
spacing for the unquenched calculation. While we would also expect
the discretization error to be small for the unquenched calculation,
this should be checked in future studies.

%% file: text_sections/dynamical.tex
%
%

\ifnum\theOutline=1
\begin{outline}
\item hadron masses and scale
\item $\alpha$ and $\beta$: $t$-dep, $m_f$-dep, and chiral limit
\item comparison with quench
\item summary table for $\alpha$ and $\beta$
\end{outline}
\fi

Quenched calculations have, in general, an uncontrolled systematic
uncertainty. 
In the previous section, we estimated the systematic error due to
quenching as approximately the size of scale ambiguity.
This is clearly an unsatisfactory technique, and full QCD calculations
are necessary.
Ultimately this can be completed by unquenching the $u$, $d$ and
$s$ quarks in the direct calculation of the form factors.
As a first step toward this direction, we shall examine the low energy
parameters $\alpha$ and $\beta$ with 
dynamical $u$ and $d$ quarks, while still treating the
s-quark in the quenched approximation.

\subsection{Description of the simulation}

We use the two-flavor dynamical DWF configurations described in \cite{Aoki:2004ht}.
These lattices were generated using the 
DBW2 gauge action with $6/g^2=0.8$,
on lattices of size $16^3\times 32$,
a fifth dimension of $L_s=12$, and  a domain wall height of
$M_5=1.8$ (see Table \ref{tab:params}).
Periodic boundary condition was imposed for all except 
the temporal direction of dynamical fermions and 
valence fermions for the spectrum and matrix elements,
where anti-periodic boundary 
conditions were used.
The dynamical quark masses are $m_f^{dyn}=0.02$, $0.03$, $0.04$,
which approximately covers the range from the half strange 
to strange quark mass.
All the analysis was carried out for 
the valence masses equal to the dynamical masses $m_f^{val}=m_f^{dyn}$.

For the mass and matrix element calculation Coulomb gauge fixed wall 
sources were used.
Measurements were performed twice per lattice: with the two source time
slices separated by the half size  
of the temporal direction $L_\tau/2$, which effectively 
doubled the statistics in comparison to the
analysis done in Ref.~\cite{Aoki:2004ht}.

\begin{figure}
 \begin{center}
  \includegraphics[width=10cm]{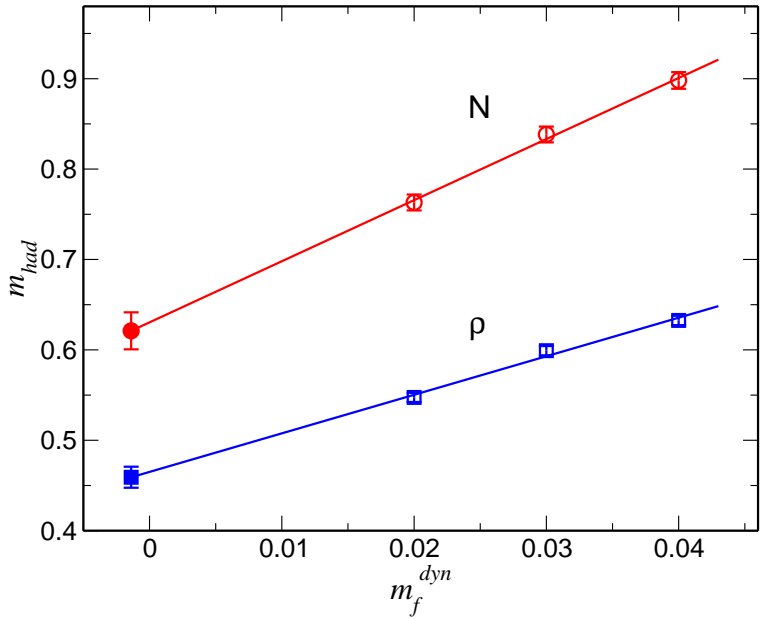}
  \caption{$\rho$ and nucleon masses as functions of $m_f^{dyn}=m_f^{val}$.}
  \label{fig:nuc_rho}
 \end{center}
\end{figure}

The results of the mass measurements are summarized in Table 
\ref{tab:masses}. A linear chiral extrapolation 
was performed for all the quantities 
in the table using all three quark masses. Figure \ref{fig:nuc_rho} 
shows $\rho$ and nucleon mass and their chiral extrapolations.
\begin{table}
 \caption{$\mres$, $Z_A$ and hadron masses from the wall source
 propagator for two-flavor dynamical domain-wall fermions with
 $6/g^2=0.8$ DBW2 gauge action.
 Two quark propagators with different source position are analyzed 
 for each configuration.
 The nucleon decay low energy parameters $\alpha$ and $\beta$ 
 for the bare  operator in lattice units are also shown.
 We have performed a linear chiral extrapolation for all 
 quantities
 using all dynamical masses $m_f=0.02$, $0.03$, $0.04$.
 The chiral limit is defined as  $m_f\to-\mres$, except $m_f\to 0$ for
 $\mres$. For $m_\pi$, squared values are fit and chiral limit shows the
 extrapolated $m_\pi^2$.}
 \label{tab:masses}
 \begin{center}
  \begin{tabular}{clllllll}
   \hline
   \hline
   \multicolumn{1}{c}{$m_f$} & \multicolumn{1}{c}{$\mres (\times 10^{-3})$}
   & \multicolumn{1}{c}{$Z_A$} & \multicolumn{1}{c}{$m_\pi$} & \multicolumn{1}{c}{$m_\rho$}
   & \multicolumn{1}{c}{$m_N$}
   & \multicolumn{1}{c}{$\alpha (\times 10^{-3})$}
   & \multicolumn{1}{c}{$\beta (\times 10^{-3})$}\\
   \hline
   \hline
   0.02 & 1.360(26) & 0.76035(27) & 0.2916(15) & 0.5474(50) & 0.7631(88) & $-$4.77(25) & 4.74(25)\\
   0.03 & 1.357(21) & 0.76187(23) & 0.3563(13) & 0.5991(51) & 0.8383(87) & $-$5.27(25) & 5.13(25)\\
   0.04 & 1.336(24) & 0.76323(21) & 0.4084(22) & 0.6325(52) & 0.8980(92) & $-$5.97(31) & 5.95(32)\\
   \hline
   chiral limit & 1.388(55) & 0.7573(6) & -0.0029(25) & 0.459(11) & 0.621(20) & $-$3.47(62) & 3.43(61)\\
   $\chi^2/dof$ & 0.10/1 & 0.08/1 & 0.55/1 & 2.11/1 & 0.51/1 & 0.11/1 & 0.42/1\\
   \hline
   \hline
  \end{tabular}
 \end{center}
\end{table}

The lattice spacing determined in the chiral limit by 
inputting $m_\rho=0.77$ GeV is
\begin{equation}
 a_\rho^{-1} = 1.678(40)\; \mbox{GeV}.
\end{equation}
which is consistent with, but, has  improved
statistical error to, the previous estimate 
gained using a
single quark propagator on each lattice 
$a_\rho^{-1} = 1.691(53)\; \mbox{GeV}$ \cite{Aoki:2004ht}.
For comparison, $a_{r_0}^{-1}=1.688(21)(^{+64}_{-04})$ is obtained
\cite{Aoki:2004ht}
with $r_0=0.5$ fm input, which is  consistent with $\rho$ input.
The lattice spacing from the $\rho$ input is $a_\rho=0.1176(28)$ fm.
Thus, the spatial size of these lattices is 1.88 fm. The 
spatial size divided by the pion Compton wave length at our 
lightest mass is $LM_\pi\simeq 4.7$.
Systematic numerical studies
\cite{Orth:2005kq,AliKhan:2003cu}, as well as theoretical
calculations \cite{Beane:2004tw}, on the finite volume effects in the nucleon
mass with two flavor Wilson fermions indicate just a few percent finite
volume mass shift with similar parameters to our lightest point. 
By this we expect the volume effect on the low energy parameters
are also small and negligible compared with the statistical error.

The ratio $m_N/m_\rho=1.353(47)$ at
the chiral limit is larger by $3\sigma$ from physical value $1.218$. 
While this discrepancy is of a size that could be 
easily ascribed to scaling violations, it should be noted that 
a similar size of discrepancy for the ratio 
has been observed in the continuum limit of the two-flavor Wilson fermion
\cite{AliKhan:2001tx} using the polynomial chiral extrapolation
for the masses. This suggests that our discrepancy might persist towards the
continuum limit. 
There is a way to cure this problem by employing the higher order
chiral expansion for the nucleon mass with terms non-analytic in the
quark mass  \cite{AliKhan:2003cu,Procura:2003ig}.

\subsection{Matrix elements}

For the non-perturbative renormalization of the operators
we follow the same procedure as for the quenched calculation.
The dynamical mass points $m_f=0.02$, $0.03$, $0.04$ are used to
analyze for the MOM scheme renormalization. Linear chiral extrapolation
in dynamical mass $m_f$ is carried out.
The resulting renormalization factors (Eq.~(\ref{eq:op_lat_to_msbar}))
which renormalize the lattice 
operator to give those in $\msbar$, NDR at $\mu=2$ GeV are
\begin{equation}
 U^{\MSbar\leftarrow latt}(2\mbox{GeV}) = \left\{
	   \begin{array}{ll}
	    0.731(28)(39) & \mbox{for } {\mathcal O}^{L;L},\cr
	    0.722(28)(39) & \mbox{for } {\mathcal O}^{R;L},
	    \end{array}
       \right.
\end{equation}
where the $\Lambda_{\MSbar}$ calculated by Alpha collaboration
for the two flavor QCD \cite{DellaMorte:2004bc},
$  \Lambda_{\MSbar}^{(2)} r_0  =  0.62(4)(4)$
has been used.

\begin{figure}[h]
 \begin{center}
  \includegraphics[width=10cm]{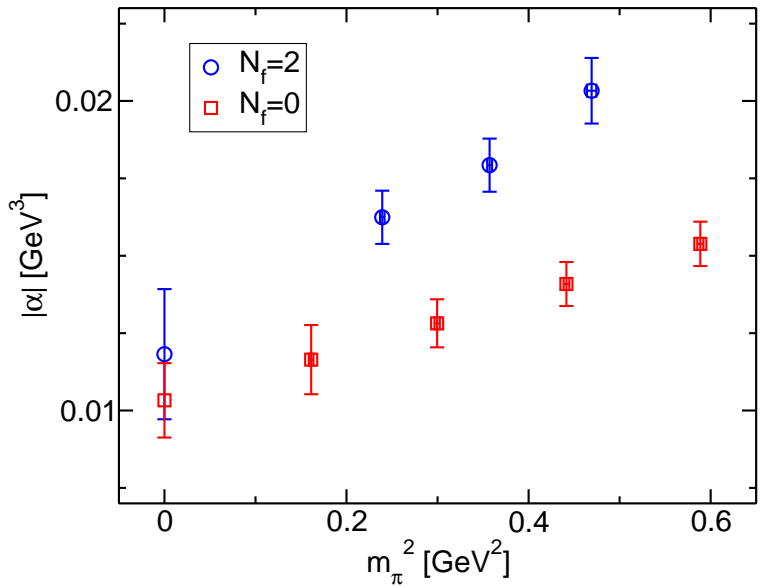}
 \end{center}
 \caption{Comparison of the quenched and dynamical $\alpha$ as a
 function of pion mass squared. Scale is set by $\rho$ input at the
 chiral limit. Operators are renormalized by $\msbar$, NDR scheme at
 $\mu=2$ GeV.}
 \label{fig:dyn_alpha}
\end{figure}

The low energy parameters are also obtained in the same way as in
the quenched case, and are shown in 
Table \ref{tab:masses}.
Figure \ref{fig:dyn_alpha} shows $\alpha$ at the
renormalization scale $\mu=2$ GeV obtained on the dynamical
configurations as a function of the pion mass squared with
scale set by $\rho$ in the chiral limit. 
The quenched results are shown for comparison.
The dynamical result has stronger $m_\pi$ dependence than quenched.
After rather long extrapolation to the chiral limit
with linear function of quark mass we obtain the $\alpha$ and $\beta$
parameters as 
\begin{eqnarray}
 -\alpha & = & 0.0118(21) \mbox{ GeV}^3,\\
 \beta & = & 0.0118(21) \mbox{ GeV}^3,
\end{eqnarray}
where errors are only statistical. Comparing to these large errors,
the errors of the renormalization constants are negligible.
These values are consistent with those obtained in quenched approximation
Eqs.~(\ref{eq:qalpha})--(\ref{eq:qbeta}).

\begin{table}
\caption{Comparison of the low energy parameter of the nucleon decay 
 chiral Lagrangian $\alpha$ and $\beta$ among various QCD model
 calculation, lattice results  in the literatures and the results from
 this work. 
 In lattice QCD
 calculations, WF and DWF mean Wilson and domain-wall fermions.
 Our quenched results are shown with the total error consisting
 of statistical and systematic errors on the bare matrix elemnt,
 renormalization constant and scale. The unquenched errors
 are only statistical.
 }
 \label{tab:summary_chL}
\begin{tabular}{cllll}
 \hline
 \hline
 & & $|\alpha|$ [GeV$^3$] & $|\beta|$ [GeV$^3$] & \\
 \hline
 \hline
 & Donoghue and Goldwich \cite{Donoghue:1982jm} & 0.003 & & Bag model\\
 & Thomas and McKellar \cite{Thomas:1983ch} & 0.02 & & Bag model\\
 & Meljanac et al. \cite{Meljanac:1982xd} & 
 0.004 & & Bag model\\
 QCD model & Ioffe \cite{Ioffe:1981kw} & 0.009 &  & Sum rule\\
 calculation & Krasnikov et al. \cite{Krasnikov:1982gf} & 
 0.003 &  & Sum rule\\
 & Ioffe and Smilga \cite{Ioffe:1984ju} & 0.006 &  & Sum rule\\
 & Tomozawa \cite{Tomozawa:1981rc} & 0.006 &  & Quark model\\
 & Brodsky et al. \cite{Brodsky:1984st} & 0.03 &  & \\
 \hline
 & Hara et al. \cite{Hara:1986hk} & 0.03 & & WF, $a=0.11$ fm\\
 & Bowler et al. \cite{Bowler:1988us} & 0.013 & 0.010 & WF, $a=0.22$ fm\\
 Lattice QCD & Gavela et al. \cite{Gavela:1989cp} & 0.0056(8) & $\simeq |\alpha|$ &
 WF, $a=0.09$ fm\\
 $N_f=0$ &JLQCD \cite{Aoki:1999tw} & 0.015(1) & 0.014(1) & WF, $a=0.09$ fm\\
 & CP-PACS \& JLQCD \cite{Tsutsui:2004qc} & 0.0090(09)($^{+5}_{-19}$) &
 0.0096(09)($^{+6}_{-20}$) & WF, continuum limit\\
 & This work & 0.0100(19) & 0.0108(21) & DWF, $a=0.15$ fm\\
 \hline
 \begin{tabular}{c}
  Lattice QCD\\
  $N_f=2$\\
 \end{tabular} & This work & 0.0118(21) & 0.0118(21) & DWF,
 $a=0.12$ fm\\
 \hline
 \hline
\end{tabular}
\end{table}
In table ~\ref{tab:summary_chL} our results on the low energy
parameters both for quenched and dynamical simulations are
compared with those obtained in the literature, which include
various QCD model calculations and the lattice
QCD efforts by several groups.
It should be remarked that by the efforts of recent calculations
including this work substantially have decreased the ambiguity
of the low energy parameters $\alpha$ and $\beta$. Their absolute
values fall in the range of the approximate value $0.01$ GeV$^3$, which 
lies in the middle of the wide range of the various QCD model
calculation result $0.003-0.03$ GeV$^3$.

%% file: text_sections/conclusions.tex
%
%

\ifnum\theOutline=1
\begin{outline}
\item we did DWF \& NPR
\item quenched matrix elements: $\alpha$, $\beta$:
 various systematic errors are small.
\item first unquenched results are consistent with quenched $\alpha$, $\beta$.
\item indirect method is poor, if we want to avoid, factor errors.
\item main results are quenched $W_0$ (direct).
\item outlook
\end{outline}
\fi

Using  domain-wall fermions and a non-perturbative renormalization
we have calculated the low energy matrix elements of the nucleon decay
whose operator would appear in the lowest order, low energy
effective Hamiltonian of any fundamental high energy theory 
that breaks baryon number and respects the symmetries of the
standard model at low energies. 

The low energy parameters $\alpha$ and $\beta$ are important fundamental
quantities of the $SU(3)_f$ baryon chiral Lagrangian with the
baryon number violating interaction. In the quenched approximation 
they were estimated at our lattice spacing ($a\simeq 1.5$ fm) as
$\alpha=-0.0100(19)$, 
$\beta=0.0108(21)$ GeV$^3$ for the operator renormalized at $\mu=2$ GeV.
We have examined the finite volume and finite lattice spacing effects
on $\alpha-\beta$, which appeared to be negligible.
These results are consistent with those using Wilson fermions 
in the continuum limit \cite{Tsutsui:2004qc}. 
The dynamical quark effects on $\alpha$ and $\beta$ were examined by
unquenching $u$ and $d$ quarks. Their values at the chiral limit
are consistent with those in the quenched approximation.
All the results so far indicate
$|\alpha|\simeq|\beta|\simeq 0.01$ GeV$^3$,
which lies in the middle of the various model calculations 
($0.003-0.03$ GeV$^3$).

However, these parameters are not quite useful for the physical
decay process. The results of the nucleon decay form factors 
calculated with $\alpha$ and $\beta$ showed clear deviation from the direct
calculation for several processes. This is presumably due to 
the large energy of the pseudoscalar meson, and because of
neglecting some leading quark mass dependent terms in the
chiral Lagrangian.
Instead, it is advised to use the 
results with the direct calculation (Table \ref{tab:W0}) for the 
partial width of proton, Eq.~(\ref{eq:width}).
The value of the form factor with the direct method tends to be smaller
compared to the indirect method; thus the direct method, which is more
accurate, tends to prolong the proton life-time. Indeed, for the case
of $p \to e^++\pi^0$, experimentally which is 
important especially for model independent analysis
such like Ref.~\cite{Dorsner:2004xa},
the difference in our central values (with appreciable error bars)
for the form factors tend to approach factor of about two. It is
clearly important to improve the precision of the direct method
in the future calculations.

The dynamical quark effects on the low energy parameters 
the mass dependence became stronger on unquenching.
However, this comparison was performed at a single lattice
spacing, and with only two -- relatively heavy -- dynamical
quarks. A clear target for the future is to repeat this
calculation with three flavors of dynamical quarks with masses near the
physical values and multiple
lattice spacings.

%% file: text_sections/appendix_notation.tex
%
%
%

\ifnum\theOutline=1
\begin{outline}
\item State normalization
\item spinor normalization
\item hadron interpolating fields
\end{outline}
\fi

In this appendix, various conventions in our calculation are
summarized.

Let $\vec{p}$ and $\vec{k}$ be spatial momenta and $i$ and $j$
symbolically denoting the other discrete quantum numbers,
the state normalization is given by
\begin{equation}
 \langle \vec{p}, i | \vec{k}, j \rangle = 
  (2\pi)^3 2 E(\vec{p},i) \delta^3(\vec{k}-\vec{p}) \delta_{i,j},
\end{equation}
where $E(\vec{p},i)$ is the energy of the $(\vec{p},i)$ state.

Spin $1/2$ wave function $u_N(k,s)$ with momentum $k$, spin $s$,
and mass $m$, that obeys the Dirac equation $(-i\ksl u_N = m u_N)$,
has a relativistic normalization,
\begin{equation}
 \overline{u}_N(k,s) u_N(k,s') = 2m \delta_{s,s'}.
  \label{eq:spin_normalization}
\end{equation}

Hadron interpolation fields, $J_{Had}$ are given as
\begin{eqnarray}
 J_{\pi^0} & = & \frac{1}{\sqrt{2}}(
  \overline{u}\gamma_5u
  - \overline{d}\gamma_5d),\\
 J_{\pi^+} & = & \overline{d}\gamma_5u,\\
 J_{\pi^-} & = & \overline{u}\gamma_5d,\\
 J_{K^0} & = & \overline{s}\gamma_5d,\\
 J_{K^+} & = & \overline{s}\gamma_5u,\\
 J_{\eta} & = & \frac{1}{\sqrt{6}}(
  \overline{u}\gamma_5u + \overline{d}\gamma_5d - 2\overline{s}\gamma_5s),\\
 J_{p} & = & \epsilon_{ijk} ( u^{iT} C \gamma_5 d^j ) u^k,\label{eqn:Jp}\\
 J_{n} & = & \epsilon_{ijk} ( u^{iT} C \gamma_5 d^j ) d^k.
\end{eqnarray}

%% file: text_sections/appendix_chpt.tex
%
%
%

\ifnum\theOutline=1

\begin{outline}
\item summary of indirect method
\item soft pion limit for all the principal matrix elements
\end{outline}
\fi

We use the tree-level chiral perturbation theory for the nucleon
decay matrix elements \cite{Claudson:1982gh,Aoki:1999tw}. 
The strong interaction chiral Lagrangian of $SU(3)_f$ octet baryons has
coupling constants: 
$D$ and $F$ describing the coupling of the nucleon to the axial current,
$a_i$, $b_i$, where $i=1$, $2$, dealing with the leading
linear quark mass dependence. The nucleon decay specific term is
proportional to $\alpha$ 
or $\beta$ that we are calculating on the lattice.

The low energy parameters $D$ and $F$ can be related to the axial charge
associated with the baryon semi-leptonic beta decay.
$D+F=g_A^{(np)}=1.27$ is the nucleon axial charge, while
$D-F=g_A^{(\Sigma^-n)}=0.33-0.34$ can be measured by 
$\Sigma^-\to n+e^-+\overline{\nu_e}$ \cite{Hsueh:1988ar,Eidelman:2004wy}.
$a_i$ are determined by the mass difference of the baryons,
and the size is $O(1)$. For nucleon decay matrix element,
it enters through the baryon masses. While $b_i$ have direct influence
to the nucleon decay amplitude, their values are not well-known.
Although they can naturally be $O(1)$, we set them to be zero in the
following.  

Here is the summary of the value of parameters we employed,
\begin{eqnarray}
 D &=& 0.8,\\
 F &=& 0.47,\\
 f & = & 0.131 {\rm GeV},\\
 m_N & = & 0.94 {\rm GeV},\\
 m_B & = & 1.15 {\rm GeV},
\end{eqnarray}
where $m_B$ is the average baryon mass $m_B\simeq M_\Sigma \simeq M_\Lambda$.

Using these constants and the approximations:
$m_{u,d}\ll m_s\ll m_{N,B}$, $-q^2\ll m_{N,B}^2$, the relevant form
factors 
$\langle PS | {\mathcal O} | N \rangle_{0} \equiv W_0$
of all the principal matrix elements are obtained as
\begin{eqnarray}
 \langle \pi^0 | (u,d)_{R} u_L | p \rangle_0 &=&
  \frac{\alpha}{\sqrt{2}f}(1+D+F), \label{eqn:indirect pi0 RL}\\
 \langle \pi^0 | (u,d)_{L} u_L | p \rangle_0 &=&
  \frac{\beta}{\sqrt{2}f} (1+D+F),\\
 \langle  K^0  | (u,s)_{R} u_L | p \rangle_0 &=&
  -\frac{\alpha}{f}\left(1+(D-F)\frac{m_N}{m_B}\right),\\
 \langle  K^0  | (u,s)_{L} u_L | p \rangle_0 &=&
  \frac{\beta}{f} \left(1-(D-F)\frac{m_N}{m_B}\right),\\
 \langle  K^+  | (u,s)_{R} d_L | p \rangle_0 &=&
  \frac{\alpha}{f}\frac{2D}{3}\frac{m_N}{m_B},\\
 \langle  K^+  | (u,s)_{L} d_L | p \rangle_0 &=&
  \frac{\beta}{f}\frac{2D}{3}\frac{m_N}{m_B},\\
 \langle  K^+  | (u,d)_{R} s_L | p \rangle_0 &=&
  \frac{\alpha}{f}\left(1+\left(\frac{D}{3}+F\right)\frac{m_N}{m_B}\right),\\
 \langle  K^+  | (u,d)_{L} s_L | p \rangle_0 &=&
  \frac{\beta}{f}\left(1+\left(\frac{D}{3}+F\right)\frac{m_N}{m_B}\right),\\
 \langle  K^+  | (d,s)_{R} u_L | p \rangle_0 &=&
  \frac{\alpha}{f}\left(1+\left(\frac{D}{3}-F\right)\frac{m_N}{m_B}\right),\\
 \langle  K^+  | (d,s)_{L} u_L | p \rangle_0 &=&
  -\frac{\beta}{f}\left(1-\left(\frac{D}{3}-F\right)\frac{m_N}{m_B}\right),\\
 \langle \eta  | (u,d)_{R} u_L | p \rangle_0 &=&
  -\frac{\alpha}{\sqrt{6}f}(1+D-3F),\\
 \langle \eta  | (u,d)_{L} u_L | p \rangle_0 &=&
  \frac{\beta}{\sqrt{6}f}(3-D+3F). \label{eqn:indirect eta LL}
\end{eqnarray}

Nucleon to pseudoscalar and nucleon to vacuum matrix elements are
related in the zero momentum limit of the pseudoscalar by soft pion
theorem,
\begin{equation}
 \lim_{p_\mu\to 0} \langle \pi^k; p_\mu | {\mathcal O} | N \rangle
  = -\frac{i}{f}\langle 0 | [Q_5^k, {\mathcal O}] | N \rangle.
  \label{eq:soft_pion}
\end{equation}
$Q_5^k$ is the axial charge having the same $SU(3)$ flavor content
as the $\pi^k$ pion. It is 
one of all the pseudoscalar states: $\pi^{0,+,-}$, $K^{0,+}$, $\eta$,
as we can consider an ideal situation, mass-less limit of all of them.

By Eq.~(\ref{eq:formfactor}), the matrix element in the soft pion limit
is written in terms of the form factors as
\begin{equation}
 \lim_{p_\mu\to 0} \langle PS; p_\mu | {\mathcal O}^L | N \rangle
  = P_{L} [ W_0 + m_N W_q] u_N.
\end{equation}
Following relations for
$\langle PS | {\mathcal O} | N \rangle_{sp} \equiv W_0 + m_N W_q$
are obtained by the soft pion theorem 
Eq.~(\ref{eq:soft_pion}) and also by $p_\mu\to 0$ limit of the
tree-level results of the chiral perturbation theory.
\begin{eqnarray}
 \langle \pi^0 | (u,d)_{R} u_L | p \rangle_{sp} &=&
  \frac{\alpha}{\sqrt{2}f},\\
 \langle \pi^0 | (u,d)_{L} u_L | p \rangle_{sp} &=&
  \frac{\beta}{\sqrt{2}f},\\
 \langle  K^0  | (u,s)_{R} u_L | p \rangle_{sp} &=&
  -\frac{\alpha}{f},\\
 \langle  K^0  | (u,s)_{L} u_L | p \rangle_{sp} &=&
  \frac{\beta}{f},\\
 \langle  K^+  | (u,s)_{R} d_L | p \rangle_{sp} &=&
  0,\\
 \langle  K^+  | (u,s)_{L} d_L | p \rangle_{sp} &=&
  0,\\
 \langle  K^+  | (u,d)_{R} s_L | p \rangle_{sp} &=&
  \frac{\alpha}{f},\\
 \langle  K^+  | (u,d)_{L} s_L | p \rangle_{sp} &=&
  \frac{\beta}{f},\\
 \langle  K^+  | (d,s)_{R} u_L | p \rangle_{sp} &=&
  \frac{\alpha}{f},\\
 \langle  K^+  | (d,s)_{L} u_L | p \rangle_{sp} &=&
  -\frac{\beta}{f},\\
 \langle \eta  | (u,d)_{R} u_L | p \rangle_{sp} &=&
  -\frac{\alpha}{\sqrt{6}f},\\
 \langle \eta  | (u,d)_{L} u_L | p \rangle_{sp} &=&
  \frac{3\beta}{\sqrt{6}f}.
\end{eqnarray}

\begin{figure}[h]
 \begin{center}
  \includegraphics[width=10cm]{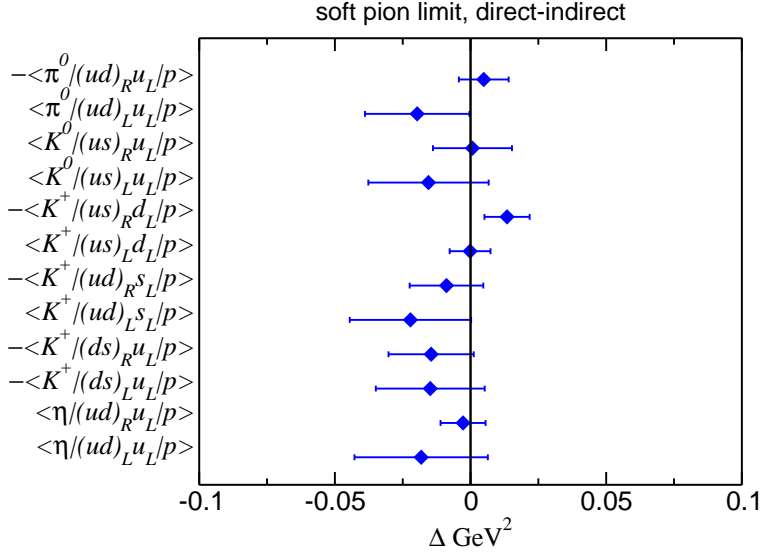}
 \end{center}
 \caption{The difference $\Delta$ of the form factors for direct and indirect 
 method in the soft pion limit. It shows the consistency with the
 soft pion theorem.
 }
 \label{fig:soft pion}
\end{figure}

These exact relations no longer involve any ordinary low energy baryonic
constants.
Checking these relations with the lattice simulation provides
a consistency test of the whole procedure.
We have measured $W_0-iq_4W_q$ with zero pseudoscalar momentum
and degenerate masses for {\it lhs}. The chiral limit should be taken with
\begin{equation}
 W_0-iq_4W_q = c_0+c_1 m_q + c_h \sqrt{m_q}.
\end{equation}
The square root of the quark mass ($\propto m_\pi$) has
entered from $q^2$ dependence. Together with $\alpha$ and $\beta$
(Eqs.~(\ref{eq:qalpha}), (\ref{eq:qbeta})) the difference
$\Delta=lhs-rhs$ is calculated and shown in Fig.~\ref{fig:soft pion}.
Most of the processes are consistent with $\Delta=0$ (note that
these are highly correlated values). Even in the worst case the
deviation is less than $2\sigma$.

%% file: text_sections/appendix_pr.tex
%
%
%

\ifnum\theOutline=1

\begin{outline}
\item MOM $to$ $\msbar$ matching.
\end{outline}
\fi

\begin{figure}[h]
  \begin{center}
  \includegraphics[width=3cm]{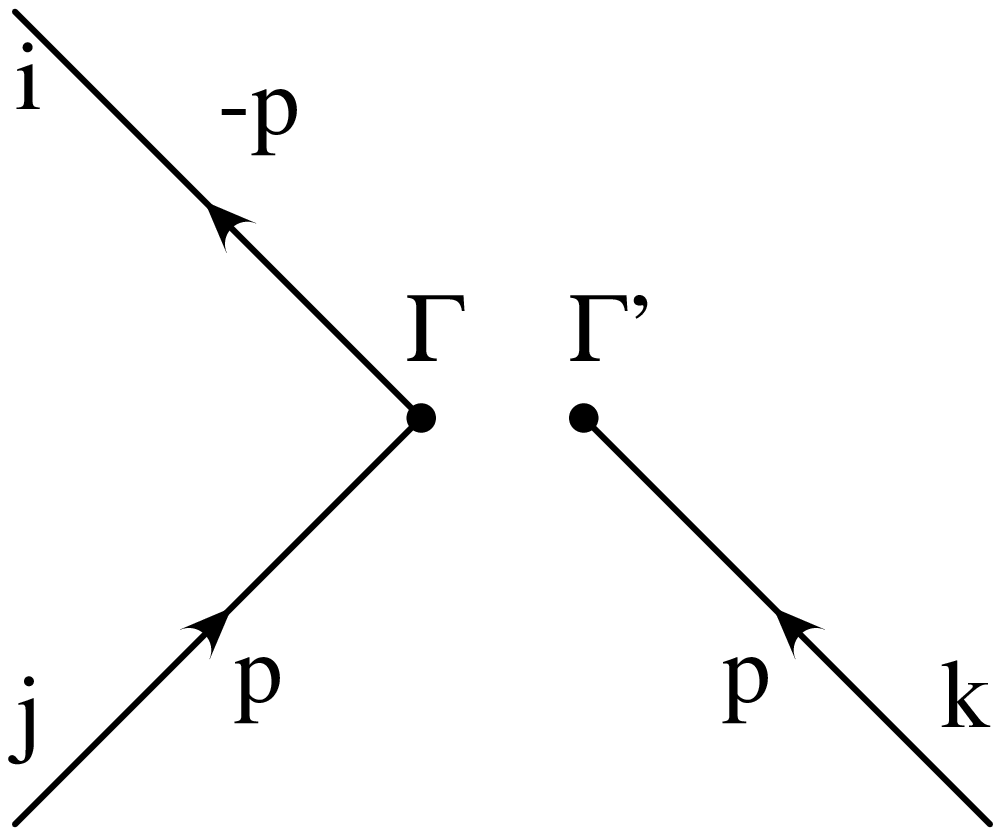}
   \mbox{\hspace{10pt}}
  \includegraphics[width=3cm]{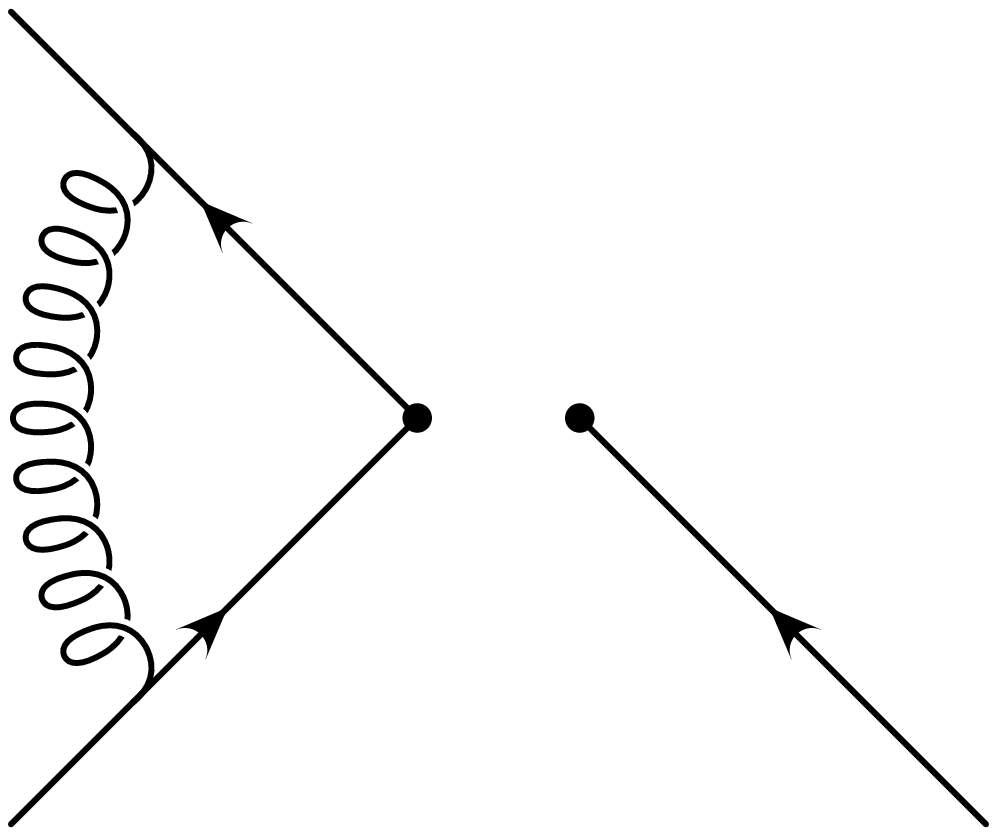}
   \mbox{\hspace{10pt}}
  \includegraphics[width=3cm]{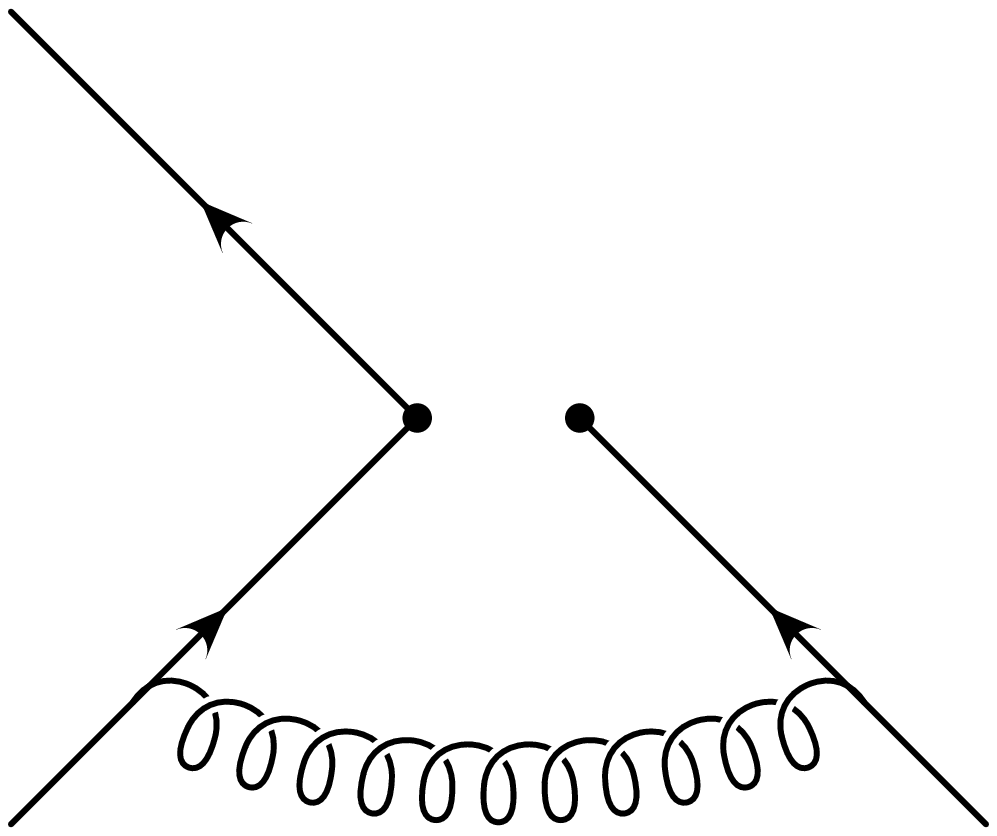}
   \mbox{\hspace{10pt}}
  \includegraphics[width=3cm]{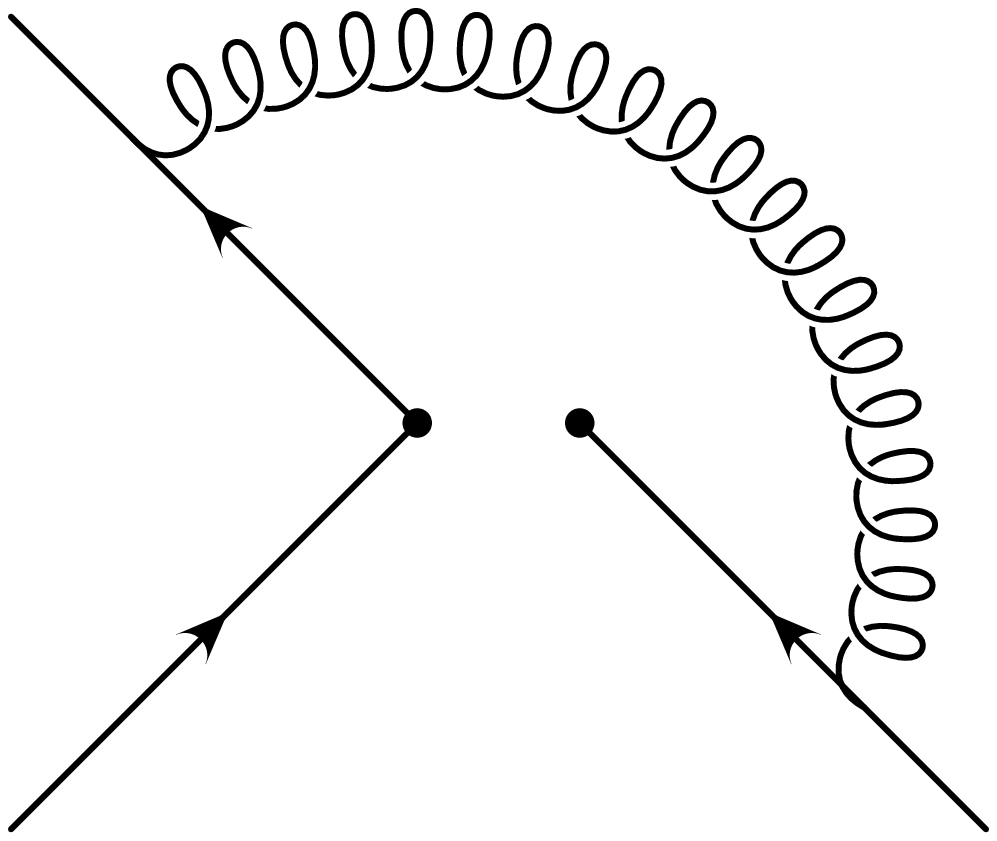}
 \end{center}
 \caption{The tree and three one-loop diagrams of the vertex function for 
 $\epsilon^{ijk}(\overline{u}^{ci} \Gamma d^j) \Gamma' s^k$.}
 \label{fig:feynman}
\end{figure}
Here we give a summary of the nucleon decay operator matching calculation.
Throughout we make use of the Minkowski space path integral.
We need to match to the lattice MOM scheme with the continuum 
$\msbar$ scheme. Since the MOM scheme is regularization independent,
we can use perturbation theory in both schemes.
In the lattice MOM scheme, the same momentum $p_\mu^E$
is injected to all the three quark external lines, where $(p^E)^2>0$. 
The corresponding Minkowski momentum is $(p^M)^2=-(p^E)^2<0$.
Both the $\msbar$ and MOM schemes are defined in the mass-less limit. We
set all masses to zero for MOM scheme in perturbation theory.

The tree-level vertex function is
\begin{equation}
 \Lambda_{ND}^{(0)} = \epsilon^{ijk} \Gamma \otimes \Gamma',
\end{equation}
where $\Gamma$ and $\Gamma'$ can be either $1$ or $\gamma_5$ or
their linear combinations. The corresponding Feynman diagram is 
shown in Fig.~\ref{fig:feynman} (leftmost). The upper left fermion line is
for the charge conjugated field, whose vertex for one gluon emission
is $-ig(t^a)^T$ whereas the normal vertex is $igt^a$.

The sum of the three one-loop diagrams in Fig.~\ref{fig:feynman} with the naive
dimensional regularization (NDR) scheme reads
\begin{equation}
 \Lambda_{ND}^{(1)} = \int\frac{d^4k}{(2\pi)^4}
  i \frac{2}{3}g^2\mu^{2\varepsilon} \epsilon^{ijk}
  [\lambda^F + (1-\xi) \lambda^{L-F}],
\end{equation}
where $\mu$ is the arbitrary mass scale introduced by the dimensional
regularization, $\varepsilon=(4-D)/2$, $D$ is the regularized dimension,
$\xi$ is the gauge parameter ($\xi=0$ is the Landau gauge). $\lambda^F$
and $\lambda^{L-F}$ read
\begin{eqnarray}
 \lambda^F & = & \frac{(D+2)(p^2-k^2)\Gamma \otimes \Gamma'}
  {[(p-k)^2+i\epsilon][(p+k)^2+i\epsilon][k^2+i\epsilon]}\\
 \lambda^{L-F} & = & \frac{3\Gamma \otimes \Gamma'}
  {[(p-k)^2+i\epsilon][(p+k)^2+i\epsilon]}\nonumber\\
 & - & \frac{p^2 \Gamma \otimes \Gamma'}
  {[(p-k)^2+i\epsilon][(p+k)^2+i\epsilon][k^2+i\epsilon]}\nonumber\\
 & - & \frac{2 (p\cdot k) \Gamma \otimes \Gamma'\psl\ksl}
  {[(p-k)^2+i\epsilon][(p+k)^2+i\epsilon][k^2+i\epsilon]^2}.
\end{eqnarray}
Carrying out the momentum integral and 
identifying $\mu^2$ with $-p^2$, one finally finds
\begin{equation}
 \Lambda_{ND}^{(1)} = \frac{\alpha_s}{4\pi}
  \left\{ [4-2(1-\xi)]\bar{\varepsilon}^{-1}
   + \left(\frac{20}{3}-16\ln2\right)
   - (1-\xi) \left(\frac{767}{180}-\frac{317}{90}\ln2\right)
  \right\} \Lambda_{ND}^{(0)},
\end{equation}
where $\bar{\varepsilon}^{-1}=2/(4-D)-\gamma_E+\ln4\pi$.
The renormalization condition of the MOM scheme up to one loop reads
\begin{equation}
 \Lambda_{ND}^{(0)} = (Z_q^{MOM})^{-3/2} Z_{ND}^{MOM} 
  (\Lambda_{ND}^{(0)}+\Lambda_{ND}^{(1)}).
\end{equation}
The MOM scheme quark wave function renormalization is given as 
(see for example \cite{Franco:1998bm})
\begin{equation}
 Z_q^{MOM} = 1 + \frac{\alpha_s}{4\pi}
  \left[\frac{4}{3}(-\xi)\bar{\varepsilon}^{-1}+\frac{2}{3}(-\xi)\right].
\end{equation}
Then the nucleon decay operator renormalization factor is calculated
as
\begin{equation}
 Z_{ND}^{MOM} = 1 + \frac{\alpha_s}{4\pi}
  \left[2\bar{\varepsilon}^{-1} + \frac{433}{180}-\frac{1123}{90}\ln 2
   + \xi \left(\frac{587}{180} - \frac{317}{90}\ln 2\right)\right].
\end{equation}
The $\msbar$ renormalization factor is given by taking only the term
proportional to $\bar{\varepsilon}^{-1}$ in the square bracket.
The matching renormalization factor $Z^{\MSbar}/Z^{MOM}$ has been
shown in Eq.~(\ref{eq:match}).

%% file: text_sections/appendix_table.tex
%
%
%

\ifnum\theOutline=1

\begin{outline}
\item table of W0 for all mass and non-zero momentum
\item W0 global fit table
\end{outline}
\fi

All the results for the relevant form factor $W_0$ of the principal matrix
elements are shown in Tables ~\ref{tab:all_W0_1}-\ref{tab:all_W0_3}.
The fit result of the relevant form factors with 
 Eq.~(\ref{eq:W_0 fit formula}) is presented in Table \ref{tab:all_fit_W0}.
\begin{table}[h]
 \caption{Relevant form factor $W_0$ of the principal matrix elements
 in lattice unit with the unrenormalized operator for each mass and
 momentum calculated on the quenched $a=0.15$ 
 fm configurations. $\vec{p}_1=(1,0,0)\cdot 2\pi/L_\sigma$ is for the
 smallest, $\vec{p}_2=(1,1,0)\cdot 2\pi/L_\sigma$ is for the second
 smallest momentum.}
 \label{tab:all_W0_1}
 \newcolumntype{.}{D{.}{.}{6}}
 \begin{tabular}{cc|....}
  \hline
  \hline
  &
  & \multicolumn{2}{c}{$RL$ operator} & \multicolumn{2}{c}{$LL$ operator}\\
  $m_1$ & $m_2$
  & \multicolumn{1}{c}{$\vec{p}=\vec{p}_1$}
  & \multicolumn{1}{c}{$\vec{p}=\vec{p}_2$}
  & \multicolumn{1}{c}{$\vec{p}=\vec{p}_1$}
  & \multicolumn{1}{c}{$\vec{p}=\vec{p}_2$}\\
  \hline
  & & \multicolumn{4}{c}{$\langle\pi^0|(ud)u|p\rangle$}\\
  0.02 & 0.02 & -0.053 (15) & 0.051 (44) & 0.061 (19) & 0.057 (72) \\
  0.02 & 0.04 & -0.060 (12) & 0.009 (21) & 0.065 (14) & 0.055 (35) \\
  0.02 & 0.06 & -0.064 (11) & -0.011 (15) & 0.067 (13) & 0.059 (24) \\
  0.02 & 0.08 & -0.066 (11) & -0.023 (13) & 0.069 (12) & 0.060 (18) \\
  0.04 & 0.04 & -0.071 (8) & -0.036 (10) & 0.075 (8) & 0.061 (14) \\
  0.04 & 0.06 & -0.073 (7) & -0.042 (8) & 0.075 (7) & 0.060 (10) \\
  0.04 & 0.08 & -0.073 (6) & -0.047 (7) & 0.074 (7) & 0.061 (8) \\
  0.06 & 0.06 & -0.082 (6) & -0.056 (7) & 0.082 (6) & 0.064 (7) \\
  0.06 & 0.08 & -0.081 (5) & -0.058 (6) & 0.080 (5) & 0.064 (6) \\
  0.08 & 0.08 & -0.088 (5) & -0.065 (5) & 0.086 (5) & 0.068 (5) \\
 \hline
  & & \multicolumn{4}{c}{$\langle K^0|(us)u|p\rangle$}\\
  0.02 & 0.02 & 0.054 (13) & 0.050 (43) & 0.035 (14) & 0.019 (49) \\
  0.02 & 0.04 & 0.056 (11) & 0.035 (22) & 0.036 (10) & 0.016 (24) \\
  0.02 & 0.06 & 0.057 (11) & 0.035 (17) & 0.039 (9) & 0.019 (17) \\
  0.02 & 0.08 & 0.059 (10) & 0.038 (14) & 0.042 (8) & 0.023 (14) \\
  0.04 & 0.04 & 0.065 (7) & 0.045 (11) & 0.031 (5) & 0.017 (10) \\
  0.04 & 0.06 & 0.066 (7) & 0.047 (9) & 0.034 (4) & 0.024 (8) \\
  0.04 & 0.08 & 0.066 (6) & 0.050 (8) & 0.037 (4) & 0.029 (6) \\
  0.06 & 0.06 & 0.069 (5) & 0.053 (7) & 0.033 (3) & 0.029 (5) \\
  0.06 & 0.08 & 0.069 (5) & 0.055 (6) & 0.036 (3) & 0.033 (4) \\
  0.08 & 0.08 & 0.072 (4) & 0.058 (5) & 0.036 (2) & 0.034 (4) \\
  \hline
  \hline
\end{tabular}
\end{table}

\begin{table}[h]
 \caption{Continued from Table \ref{tab:all_W0_1}.}
 \label{tab:all_W0_2}
 \newcolumntype{.}{D{.}{.}{6}}
 \begin{tabular}{cc|....}
  \hline
  \hline
  &
  & \multicolumn{2}{c}{$RL$ operator} & \multicolumn{2}{c}{$LL$ operator}\\
  $m_1$ & $m_2$
  & \multicolumn{1}{c}{$\vec{p}=\vec{p}_1$}
  & \multicolumn{1}{c}{$\vec{p}=\vec{p}_2$}
  & \multicolumn{1}{c}{$\vec{p}=\vec{p}_1$}
  & \multicolumn{1}{c}{$\vec{p}=\vec{p}_2$}\\
  \hline
  & & \multicolumn{4}{c}{$\langle K^+|(us)d|p\rangle$}\\
  0.02 & 0.02 & -0.020 (9) & 0.013 (29) & 0.017 (10) & 0.017 (35) \\
  0.02 & 0.04 & -0.020 (7) & 0.002 (13) & 0.018 (7) & 0.015 (16) \\
  0.02 & 0.06 & -0.020 (6) & -0.001 (9) & 0.019 (6) & 0.016 (11) \\
  0.02 & 0.08 & -0.019 (5) & -0.003 (8) & 0.018 (5) & 0.017 (8) \\
  0.04 & 0.04 & -0.029 (4) & -0.013 (6) & 0.025 (3) & 0.021 (6) \\
  0.04 & 0.06 & -0.028 (4) & -0.014 (4) & 0.024 (3) & 0.018 (4) \\
  0.04 & 0.08 & -0.027 (3) & -0.014 (4) & 0.023 (3) & 0.017 (3) \\
  0.06 & 0.06 & -0.033 (3) & -0.019 (3) & 0.027 (2) & 0.019 (3) \\
  0.06 & 0.08 & -0.031 (3) & -0.019 (3) & 0.026 (2) & 0.018 (3) \\
  0.08 & 0.08 & -0.035 (2) & -0.023 (2) & 0.029 (2) & 0.020 (2) \\
 \hline
  & & \multicolumn{4}{c}{$\langle K^+|(ud)s|p\rangle$}\\
  0.02 & 0.02 & -0.056 (17) & 0.067 (48) & 0.068 (19) & 0.073 (74) \\
  0.02 & 0.04 & -0.065 (14) & 0.011 (22) & 0.073 (15) & 0.067 (37) \\
  0.02 & 0.06 & -0.070 (13) & -0.014 (16) & 0.076 (13) & 0.069 (25) \\
  0.02 & 0.08 & -0.074 (12) & -0.031 (14) & 0.079 (12) & 0.070 (20) \\
  0.04 & 0.04 & -0.072 (8) & -0.038 (11) & 0.081 (8) & 0.066 (15) \\
  0.04 & 0.06 & -0.075 (7) & -0.047 (9) & 0.082 (7) & 0.067 (11) \\
  0.04 & 0.08 & -0.077 (7) & -0.053 (8) & 0.083 (7) & 0.069 (9) \\
  0.06 & 0.06 & -0.082 (6) & -0.060 (7) & 0.088 (6) & 0.071 (7) \\
  0.06 & 0.08 & -0.083 (6) & -0.063 (7) & 0.088 (6) & 0.072 (7) \\
  0.08 & 0.08 & -0.090 (5) & -0.069 (6) & 0.093 (5) & 0.076 (6) \\
  \hline
  \hline
 \end{tabular}
\end{table}

\begin{table}[h]
 \caption{Continued from Table \ref{tab:all_W0_2}.}
 \label{tab:all_W0_3}
 \newcolumntype{.}{D{.}{.}{6}}
 \begin{tabular}{cc|....}
  \hline
  \hline
  &
  & \multicolumn{2}{c}{$RL$ operator} & \multicolumn{2}{c}{$LL$ operator}\\
  $m_1$ & $m_2$
  & \multicolumn{1}{c}{$\vec{p}=\vec{p}_1$}
  & \multicolumn{1}{c}{$\vec{p}=\vec{p}_2$}
  & \multicolumn{1}{c}{$\vec{p}=\vec{p}_1$}
  & \multicolumn{1}{c}{$\vec{p}=\vec{p}_2$}\\
  \hline
  & & \multicolumn{4}{c}{$\langle K^+|(ds)u|p\rangle$}\\
  0.02 & 0.02 & -0.034 (9) & -0.058 (43) & -0.051 (14) & -0.065 (54) \\
  0.02 & 0.04 & -0.035 (7) & -0.038 (22) & -0.054 (11) & -0.050 (27) \\
  0.02 & 0.06 & -0.037 (7) & -0.035 (16) & -0.057 (9) & -0.050 (19) \\
  0.02 & 0.08 & -0.039 (6) & -0.037 (13) & -0.060 (9) & -0.051 (15) \\
  0.04 & 0.04 & -0.035 (5) & -0.033 (10) & -0.056 (6) & -0.043 (11) \\
  0.04 & 0.06 & -0.037 (4) & -0.034 (8) & -0.058 (5) & -0.047 (8) \\
  0.04 & 0.08 & -0.039 (4) & -0.036 (7) & -0.060 (5) & -0.050 (7) \\
  0.06 & 0.06 & -0.036 (3) & -0.033 (5) & -0.061 (4) & -0.050 (6) \\
  0.06 & 0.08 & -0.038 (3) & -0.036 (5) & -0.062 (4) & -0.053 (5) \\
  0.08 & 0.08 & -0.037 (2) & -0.035 (4) & -0.065 (4) & -0.055 (4) \\
 \hline
  & & \multicolumn{4}{c}{$\langle\eta|(ud)u|p\rangle$}\\
  0.02 & 0.02 & 0.013 (7) & 0.072 (42) & 0.062 (17) & 0.079 (66) \\
  0.02 & 0.04 & 0.011 (5) & 0.035 (19) & 0.066 (13) & 0.061 (32) \\
  0.02 & 0.06 & 0.010 (5) & 0.023 (13) & 0.070 (11) & 0.061 (23) \\
  0.02 & 0.08 & 0.009 (5) & 0.018 (11) & 0.073 (11) & 0.063 (18) \\
  0.04 & 0.04 & 0.010 (3) & 0.016 (8) & 0.068 (7) & 0.053 (13) \\
  0.04 & 0.06 & 0.010 (3) & 0.014 (6) & 0.071 (6) & 0.057 (10) \\
  0.04 & 0.08 & 0.011 (2) & 0.013 (5) & 0.074 (6) & 0.061 (8) \\
  0.06 & 0.06 & 0.009 (2) & 0.010 (4) & 0.075 (5) & 0.061 (7) \\
  0.06 & 0.08 & 0.009 (2) & 0.011 (3) & 0.076 (5) & 0.065 (6) \\
  0.08 & 0.08 & 0.008 (1) & 0.009 (2) & 0.079 (4) & 0.067 (5) \\
  \hline
  \hline
 \end{tabular}
\end{table}

\begin{table}[h]
 \caption{Fit results of the relevant form factor $W_0$ with 
 Eq.~(\ref{eq:W_0 fit formula}) for the
 principal matrix elements. Shown values are in lattice unit and
 by unrenormalized operators. $dof=16$ for all.}
 \label{tab:all_fit_W0}
 \newcolumntype{.}{D{.}{.}{6}}
 \begin{tabular}{c....l}
  \hline
  \hline
  & \multicolumn{1}{c}{$c_0$}
  & \multicolumn{1}{c}{$c_1$}
  & \multicolumn{1}{c}{$c_2$}
  & \multicolumn{1}{c}{$c_3$}
  & \multicolumn{1}{c}{$\chi^2$} \\
  \hline
$\langle\pi^0|(ud)_Ru_L|p\rangle$ & -0.047 (13) & -0.125 (31) & -0.24 (13) & -0.27 (8) & 14 (11) \\
$\langle\pi^0|(ud)_Lu_L|p\rangle$ & 0.067 (15) & 0.070 (33) & 0.09 (16) & 0.09 (10) & 0.6 (22) \\
$\langle K^0|(us)_Ru_L|p\rangle$ & 0.055 (12) & 0.071 (32) & 0.07 (14) & 0.12 (7) & 1.6 (24) \\
$\langle K^0|(us)_Lu_L|p\rangle$ & 0.026 (9) & 0.023 (23) & -0.07 (10) & 0.19 (6) & 4.5 (61) \\
$\langle K^+|(us)_Rd_L|p\rangle$ & -0.021 (7) & -0.057 (16) & -0.13 (7) & -0.02 (5) & 4.7 (55) \\
$\langle K^+|(us)_Ld_L|p\rangle$ & 0.023 (6) & 0.030 (14) & 0.07 (7) & -0.02 (4) & 1.9 (39) \\
$\langle K^+|(ud)_Rs_L|p\rangle$ & -0.046 (14) & -0.116 (34) & -0.18 (15) & -0.35 (9) & 15 (12) \\
$\langle K^+|(ud)_Ls_L|p\rangle$ & 0.072 (15) & 0.069 (35) & 0.06 (16) & 0.15 (11) & 0.4 (19) \\
$\langle K^+|(ds)_Ru_L|p\rangle$ & -0.033 (8) & -0.010 (22) & 0.06 (9) & -0.10 (5) & 0.5 (16) \\
$\langle K^+|(ds)_Lu_L|p\rangle$ & -0.049 (11) & -0.043 (26) & 0.01 (12) & -0.17 (7) & 0.5 (10) \\
$\langle\eta|(ud)_Ru_L|p\rangle$ & 0.013 (6) & -0.012 (15) & -0.04 (6) & -0.02 (4) & 4.7 (66) \\
$\langle\eta|(ud)_Lu_L|p\rangle$ & 0.061 (13) & 0.053 (32) & -0.01 (14) & 0.20 (9) & 0.5 (10) \\
\hline
\hline
\end{tabular}
\end{table}